\documentclass[12pt]{article}

\usepackage[margin=1in]{geometry}
\usepackage{setspace}
\usepackage{float}
\usepackage{appendix}
\usepackage{titletoc}

\usepackage[T1]{fontenc}
\usepackage[utf8]{inputenc}
\usepackage{soul}

\usepackage{amsmath,amssymb,amsfonts,amsthm,bbm}
\usepackage{mathtools}

\usepackage{graphicx}
\usepackage{booktabs}
\usepackage{caption}
\usepackage{subcaption}
\usepackage{pbox}

\usepackage{natbib}
\usepackage{hyperref}
\usepackage{xcolor}
\hypersetup{
    colorlinks=true,
    breaklinks=true,
    bookmarks=true,
    bookmarksnumbered=true,
    urlcolor=red,
    linkcolor=red,
    citecolor=blue
}

\usepackage[linesnumbered,ruled,vlined]{algorithm2e}

\newtheorem{assumption}{Assumption}[section]

\newtheorem{corollary}{Corollary}[section]
\newtheorem{proposition}{Proposition}[section]
\newtheorem*{proposition_no}{Proposition}
\newtheorem{remark}{Remark}[section]

\usepackage{xr}
\begin{document}

  \doublespacing

\title{ {
  Inference in Difference-in-Differences with Few Treated Units and Spatial Correlation}}

\author{
Luis A. F. Alvarez\thanks{email: luis.alvarez@usp.br; address: Department of Economics, University of São Paulo, Luciano Gualberto 908, São Paulo, Brazil, 05508-010; telephone number: +55 11 3091-5972.}  \and Bruno Ferman\thanks{email: bruno.ferman@fgv.br; address: São Paulo School of Economics, FGV, Rua Itapeva 474, São Paulo, Brazil, 01332-000; telephone number: +55 11 3799-3350.} 
}

\maketitle

	
{
\singlespacing		

\begin{abstract}
    
\noindent
We consider the problem of inference in Difference-in-Differences (DID)  when there are few treated units and errors are spatially correlated. We first show that, when there is a single  treated unit, some existing inference methods designed for settings with few treated and many control units remain asymptotically valid when errors are weakly dependent. However, these methods may be invalid with more than one treated unit. We propose a menu of  alternatives that are asymptotically valid in this setting, even when the relevant distance metric across units is unavailable. These alternatives vary in terms of the length of the resulting confidence intervals and the strength of the required assumptions. Our methods are also valid for comparison-of-means estimators and for construction of prediction intervals for counterfactual imputation methods.

\vspace{1em}  

\noindent%

\vspace{1em}  

\end{abstract}

}

\doublespacing

\section{Introduction}

Difference-in-Differences (DID) presents several challenges for inference, and a wide range of methods has been developed to address them. The effectiveness of each approach depends critically on the assumptions imposed on the errors and on features of the empirical design, such as the number of treated and control units. A non-exhaustive list of papers that proposed or analyzed different inference methods for DID in different settings include \cite{Arellano}, \cite{Bertrand04howmuch}, \cite{Donald}, \cite{cameron2008bootstrap}, \cite{CT}, \cite{Bester2011}, \cite{Muller2016}, \cite{IZA}, \cite{Canay}, \cite{FP}, \cite{MW}, \cite{ferman2019simple,FermanJAE}, \cite{Roth2020}, \cite{Athey2022}, and \cite{alvarez2023extensions}.

We focus on a common setting for which a satisfactory solution is not yet available: (i) there is a small number of treated units, (ii) the number of periods is fixed, and (iii) errors may be correlated across units, but the relevant distance metric is unavailable to the researcher. Throughout, we use ``spatial correlation''  broadly to mean any cross-sectional correlation, not necessarily tied to geography. Our results are also valid for cross-section comparison of means with few treated and many control units, and also for the construction of prediction intervals for counterfactual imputation methods \citep{AF2025}. 

\cite{CT} (henceforth CT) and \cite{FP} (henceforth FP) propose inference methods for cases with few treated and many control units when the number of pre-treatment periods is fixed. Their validity, however, relies on either independence across units or spatial correlation structured by an observed distance metric. We first consider the case of a single treated unit and derive conditions under which these methods remain asymptotically valid even with spatial correlation, when no distance metric is available. The main assumptions are: (i) the post-pre difference in average errors has the same marginal distribution across units (relaxable to allow for heteroskedasticity with a known estimable structure); (ii) there is no stochastic treatment effect heterogeneity (though we also discuss interpretations following \cite{AF2025} when this assumption fails); and (iii) the cross-sectional distribution of the post-pre difference in average errors is weakly dependent among control units. Under these conditions, the asymptotic distribution of the DID estimator depends only on the treated unit's post-pre error difference, and the residuals of the control units asymptotically recover the treated unit’s error distribution---even with spatial correlation.

When more than one treated unit is present, however, the CT and FP methods may fail to be asymptotically valid under spatial correlation,  even if the assumptions above are satisfied. The intuition is straightforward: assuming independence across treated clusters understates the variability of their average when errors are positively correlated, implying that we may over-reject the null. We therefore propose a menu of inference procedures that remain asymptotically valid --- though generally conservative --- in the presence of spatial correlation.

We start considering methods that are valid without imposing any restriction on the spatial correlation. We use the results that (i) the marginal distributions of the post-pre errors of each treated unit is identified and (ii) the asymptotic distribution of the DID estimator depends only on the average of the post-pre errors of the treated units. {Using tight quantile bounds under arbitrary dependence, we are able to bound} the quantiles of the asymptotic distribution of the DID estimator without imposing any assumption on the spatial correlation. By imposing positive dependence on the errors among treated units, we consider a second alternative. Since CT and FP remain valid inference methods when there is a single treated unit, the main idea is to consider inference for each treated unit separately, and control for False Discovery Rate (FDR) using \cite{Benjamini1995} procedure. By inverting this multiple hypothesis testing (MHT) procedure, one can obtain valid confidence sets for heterogeneous effects and project them to construct confidence sets for the average effect. In a third alternative, we derive critical values assuming  comonotonicity among the treated units constitutes the worst-case scenario for spatial dependence. We show that this approach is conservative under an assumption on the tail of the average of the errors of the treated units, for which we provide sufficient conditions for its validity. Finally, a fourth approach is available when the outcome aggregates individual-level unit $\times$ time data, where we exploit information on the within-unit correlation to bound the between-unit correlation. Assuming individuals within the same unit are more correlated than those across units, we construct a test that is less conservative than the previous one.

We evaluate these alternatives through Monte Carlo simulations calibrated to the correlation structure of the American Community Survey (ACS), and revisit the analysis of the Massachusetts 2006 health reform in \cite{Sommers} in light of our results.  The main message from these empirical exercises is that we are able to conduct valid inference even when we consider settings with spatial correlation with an unknown distance metric. Moreover, these exercises illustrate how assuming even mild and arguably plausible assumptions can go a long way in improving the length of confidence intervals in this setting.    

Overall, we provide a menu of alternative inference methods that differ with respect to their required assumptions and the length of the confidence intervals they generate. While these approaches are valid for constructing of confidence intervals sets when treatment effects are non-stochastic, when we consider settings with stochastic treatment effects, they are also   valid for construction of prediction intervals sets, for testing sharp null hypotheses, and, under additional assumptions, also for inference on the realized treatment effects, following \cite{AF2025}.

\section{Setting}  \label{setting}

Consider a panel with $N$ units and $T$ periods. We focus on the case in which treatment is non-reversible and starts for all treated units after date $t^\ast$.\footnote{We consider the case with variation in treatment timing in Appendix \ref{Appendix_staggered}.} Therefore, we can define potential outcomes $Y_{st}(0)$ and $Y_{st}(1)$ as the potential outcomes of unit $s$ at time $t$ when this unit is untreated and treated at this period. We consider that potential outcomes are given by
\begin{eqnarray} \label{simple_did_equation}
\begin{cases} Y_{st}(0) =   \theta_s + \gamma_t + \eta_{st}  \\  Y_{st}(1) = \alpha_{st} +Y_{st}(0),  \end{cases} 
\end{eqnarray}
where $\alpha_{st}$ is the  treatment effects for unit $s$ at time $t$, $\theta_s$ are time-invariant unobserved effects, and $\gamma_t$ are group-invariant unobserved effects. The error term $\eta_{st}$ represents unobserved determinants of $Y_{st}(0)$ that are not captured by the fixed effects. We observe $Y_{st} = d_{st} Y_{st}(1) + (1-d_{st}) Y_{st}(0)$, where $d_{st}$ is a dummy variable equal to one if unit $s$ is treated at time $t$. We can also consider the case in which we observe individual-level observations $Y_{ist}$. In this case, a solution to take within-(unit $\times$ time) correlations into account is to consider unit $\times$ time aggregates $Y_{st}$. Therefore, we focus on the unit $\times$ time aggregate setting. 

We consider a model-based setting in which treatment assignment is fixed, so uncertainty comes from unobserved shocks that may affect the potential outcomes of unit $s$ at time $t$, such as, for example, weather or economic shocks that are unobserved by the econometrician (denoted by $\eta_{st}$), which are allowed to be serially and also spatially correlated.  See \cite{alvarez2025inferencetreatedunits} for further discussion on the use of model-based designs in settings with few treated units. Let $\mathcal{I}_1$ ($\mathcal{I}_0$) be the set of indices for treated (control) units, while  $\mathcal{T}_1$ ($\mathcal{T}_0$) be the set of indices for post- (pre-) treatment periods. Also, let $N_d =|\mathcal{I}_d|$ and $T_d=|\mathcal{T}_d|$, for $d \in \{0,1\}$.

While $\alpha_{st}$ can vary across both $s$ and $t$, we treat $\{ \alpha_{st} \}_{s \in \mathcal{I}_1,t \in \mathcal{T}_1}$ as fixed parameters, and define $\alpha \equiv \frac{1}{N_1} \frac{1}{T-t^\ast} \sum_{s \in \mathcal{I}_1}\sum_{t \in \mathcal{T}_1} \alpha_{st}$, which is the average treatment effects across treated units and treated periods (ATT). In the terminology of \cite{alvarez2025inferencetreatedunits}, this means that we allow for \emph{deterministic} treatment effect heterogeneity, but we do not allow for \emph{stochastic} treatment effect heterogeneity.\footnote{A framework in which treatment assignment and treatment effects are treated as fixed is common in the literature of DID with few treated clusters, as considered by CT, FP, and \cite{alvarez2023extensions}. A similar framework is also considered in other settings in which the number of treated clusters is fixed, such as in the synthetic controls literature \citep{Abadie2010,SDID, ASC,BF,CWZ,FP_QE,Ferman_JASA,Wang}. 
} In Remark \ref{Remark_AF} we discuss alternative interpretations of the inference methods we propose when we allow for stochastic treatment effect heterogeneity.  Since $\theta_s$ and $\gamma_t$ are eliminated when we include the fixed effects, we do not need to impose any constrain on those variables, which can be treated either as fixed or stochastic.

Let $W_s = \frac{1}{T-t^\ast} \sum_{t \in \mathcal{T}_1} \eta_{st} -  \frac{1}{t^\ast} \sum_{t \in \mathcal{T}_0} \eta_{st}$,  which is the post-pre difference in average errors for each unit $s$.   In this  case in which treatment starts at the same period for all treated units, the DID estimator is numerically equivalent to the two-way fixed effects (TWFE) estimator,  which is given by
\begin{equation}  \label{alpha}
	\begin{aligned}
\widehat \alpha  &= \frac{1}{N_1} \sum_{s \in \mathcal{I}_1} \left[ \frac{1}{T-t^\ast} \sum_{t \in \mathcal{T}_1} Y_{st} -  \frac{1}{t^\ast} \sum_{t \in \mathcal{T}_0} Y_{st} \right]   - \frac{1}{N_0} \sum_{s \in \mathcal{I}_0}  \left[ \frac{1}{T-t^\ast} \sum_{t \in \mathcal{T}_1} Y_{st} -  \frac{1}{t^\ast} \sum_{t \in \mathcal{T}_0} Y_{st} \right] \\ &= \alpha +  \frac{1}{N_1} \sum_{s \in \mathcal{I}_1} W_s - \frac{1}{N_0} \sum_{s \in \mathcal{I}_0} W_s.
\end{aligned}
\end{equation}

We consider the following identification assumption.

\begin{assumption}{} \label{As_id}
\normalfont
 $ \mathbb{E} \left[ W_s   \right] =0$ for all $s \in \mathcal{I}_1 \cup \mathcal{I}_0$.
\end{assumption}

We recall that treatment assignment is fixed, so this assumption means that the post-pre difference in average errors has the same mean for both treated and control units. This assumption is implied by a standard parallel trends assumption for all periods, and is equivalent to assuming a parallel trends assumption on the potential outcomes.  We can extend our results to consider alternative parallel trends assumptions  \citep{Marcus} and alternative estimators. 

Given Assumption \ref{As_id}, we have that  $\mathbb{E}[\widehat \alpha] = \alpha$, so the DID estimator is unbiased, regardless of whether or not errors are spatially correlated. However, in a setting in which $N_1$ is fixed and $N_0 \rightarrow \infty$, the DID estimator will not be consistent,  and will not  necessarily  be asymptotically normal. In particular, CT show that, under a strong mixing condition on the errors, in this setting $\widehat \alpha$ converges in probability to $\alpha + \widetilde W$, where $\widetilde W \equiv \frac{1}{N_1}  \sum_{s \in \mathcal{I}_1} W_s$. Therefore, if we want to test the null  hypothesis $H_0: \alpha = \alpha_0$ at the significance level $\tau$, this would pose some challenges for inference.

\begin{remark}
The main idea of CT and FP is to collapse the time series by considering post-pre differences in the average errors for each unit. Therefore, all our results are also valid for comparison-of-means settings. More generally, all results can also be extrapolated to   the construction of prediction intervals for counterfactual imputation methods \citep{AF2025}. 
\end{remark}

\section{Inference with independent clusters}
\label{Sec_independent}

Before we move to the case in which errors may be spatially correlated, we start with a brief review of inference methods for settings with few treated clusters, when errors are independent in the cross-section. We focus on the inference approaches proposed by CT and FP for settings with few treated units and a fixed number of periods.  See \cite{alvarez2025inferencetreatedunits} for a more extensive survey on inference methods for settings with few treated units.

CT propose an interesting inference method in this setting by noting that the residuals $\widehat W_s$ of the control units may be informative about the distribution of $W_s$ for the treated. In their running model, they assume that $W_s$ is iid across $s \in \mathcal{I}_1 \cup \mathcal{I}_0$ (Assumption 2 from CT). Note that, under this iid assumption, knowledge about the marginal distribution of $W_s$ for the control units implies knowledge about the distribution of $\widetilde W$. Therefore, the main idea from CT is to use $\{\widehat W_s \}_{s \in \mathcal{I}_0}$ to approximate the marginal distribution of $W_s$ for the treated, and then use that to calculate critical values.  More specifically, they propose Algorithm \ref{alg_ct}  for testing the null $H_0: \alpha = \alpha_0$ at a significance level $\tau$. 

{\singlespacing
\renewcommand{\thealgocf}{CT}
\begin{algorithm}[h]
	 Run the DID estimator, and store $\widehat \alpha$ and the residuals $\widehat W_s = \frac{1}{T-t^\ast} \sum_{t \in \mathcal{T}_1}\hat \eta_{st} -  \frac{1}{t^\ast} \sum_{t \in \mathcal{T}_0} \hat \eta_{st}$ for all $s \in \mathcal{I}_0$ \;
	 \For{b=1, \ldots, B}	{
	 Sample with replacement $N_1$ values from  $\{\widehat W_s \}_{s \in \mathcal{I}_0}$, $(\widehat W_1^\ast,...,\widehat W_{N_1}^\ast)$ \;
	  Calculate $\widehat \alpha_b = \frac{1}{N_1} \sum_{s=1}^{N_1} \widehat W_s^\ast $ \;
	}
 Reject the null if $\frac{1}{B}\sum_{b=1}^B \mathbf{1}\{  \widehat \alpha - \alpha_0  > \widehat \alpha_b  \} < \tau/2$ or  $\frac{1}{B}\sum_{b=1}^B \mathbf{1}\{  \widehat \alpha - \alpha_0  < \widehat \alpha_b  \} > 1- \tau/2$.
	 \caption{\cite{CT}}
	 \label{alg_ct}
\end{algorithm}
}

Let $\phi_{\mbox{\tiny CT}}$ be an indicator variable equal to one if we reject the null in Algorithm \ref{alg_ct}. We summarize the results from CT in the following proposition.

\begin{proposition_no}[CT]

Suppose we have data on $\{Y_{s1},...,Y_{sT}\}_{ \mathcal{I}_1 \cup \mathcal{I}_0}$, and that Assumption \ref{As_id} holds. Assume also that $W_s$ is iid across $s$, with a common distribution function $F_W$ that has bounded second moment and is absolutely continuous with bounded density. Then, as $N_0 \rightarrow \infty$ and $N_1$ is fixed, (i) $\widehat \alpha$ converges in probability to $\alpha + \widetilde{W}$, and (ii) if the null is true, then  $\lim_{N_0,B \to \infty }\mathbb{E}[\phi_{\mbox{\tiny CT}}] = \tau$.

\end{proposition_no}

Note that CT provide a proof of result (i) under less restrictive conditions (their Proposition 1).  Proposition 2 from CT presents result (ii) under similar conditions as we consider here. In particular, assuming that errors are iid in the cross section.  CT also consider another alternative for inference in their appendix, in which they relax the iid assumption, allowing for spatial correlation and heteroskedasticity with a known structure. However, this alternative relies on a known distance metric. It also requires parametrization/estimation of the serial correlation structure, and relies on normality. 

FP builds on CT to propose an alternative that allows for heteroskedasticity with a known structure that can be estimated, without requiring parametrization/estimation of the serial correlation structure, and without relying on normality. They consider a setting in which we also observe a vector of covariates $Z_s$, and assume that $W_s = h(Z_s,\delta) \xi_s$, where $h(\cdot,\delta)$ is a known function with $\delta$ being an unknown parameter, and $\xi_s$ is iid for all $s$. In Appendix \ref{app_plausible_het}, we provide evidence that, under the assumption that heteroskedasticity is a sole function of $Z_s$, our adopted parametric form for $h$ is reasonable for the dataset used to base our simulations in Section \ref{simulations}, and for the dataset of our empirical illustration in Section \ref{illustration}.  Throughout, we consider that the sequence $\{Z_s\}_{ \mathcal{I}_1 \cup \mathcal{I}_0}$ is fixed, allowing $Z_s$ to be arbitrarily different between treated and control units. This allows for heteroskedasticity with a known structure (up to a parameter that can be estimated), but still relies on independence across units. In this case, instead of directly sampling from $\{\widehat W_s \}_{s \in \mathcal{I}_0}$, we re-scale the residuals taking into account that they may have different variances. {Algorithm \ref{alg_fp} summarizes the steps to implement their inference method when testing the null $H_0: \alpha = \alpha_0$ at a significance level $\tau$.}

{\singlespacing
\renewcommand{\thealgocf}{FP}
\begin{algorithm}[h]
Run the DID estimator, and store $\widehat \alpha$ and the residuals  $\widehat W_s = \frac{1}{T-t^\ast} \sum_{t \in \mathcal{T}_1}\hat \eta_{st} -  \frac{1}{t^\ast} \sum_{t \in \mathcal{T}_0} \hat \eta_{st}$ for all $s \in \mathcal{I}_0$ \;
Estimate $\delta$ using the residuals from the controls (see example below) \;
Compute the normalized residuals $\widehat \xi_s = \widehat W_s / h(Z_s,\widehat \delta)$, for $s \in \mathcal{I}_0$ \;
	\For{b=1, \ldots, B}	{
		Sample with replacement $N_1$ values from  $\{\widehat \xi_s \}_{s \in \mathcal{I}_0}$, $(\widehat \xi_1^\ast,...,\widehat \xi_{N_1}^\ast)$ \; 
		
		Calculate $\widehat \alpha_b = \frac{1}{N_1} \sum_{s=1}^{N_1} h(Z_s,\widehat \delta) \widehat \xi_s^\ast $ \;
	}
 Reject the null if $\frac{1}{B}\sum_{b=1}^B \mathbf{1}\{  \widehat \alpha - \alpha_0  > \widehat \alpha_b  \} < \tau/2$ or  $\frac{1}{B}\sum_{b=1}^B \mathbf{1}\{  \widehat \alpha - \alpha_0  < \widehat \alpha_b  \} > 1- \tau/2$.
	\caption{\cite{FP}}
	\label{alg_fp}
\end{algorithm}
}

Let $\phi_{\mbox{\tiny FP}}$ be an indicator variable equal to one if we reject the null in Algorithm \ref{alg_fp}. 

\begin{proposition_no}[FP]

Suppose we have data on $\{Y_{s1},...,Y_{sT},Z_s\}_{ \mathcal{I}_1 \cup \mathcal{I}_0}$, and that Assumption \ref{As_id} holds. Assume also that  $W_s = h(Z_s,\delta) \xi_s$, where $\xi_s$ is iid across $s$, with a common distribution function $F_\xi$ that has bounded second moment and is absolutely continuous with bounded density. Assume there exist constants $0 <\underline{h}\leq \overline{h} <\infty$, not depending on $N_0$, such that $ \underline h \leq h(Z_s,\delta) \leq \overline{h} $ for all $s \in \mathcal{I}_0 \cup \mathcal{I}_1$, uniformly as $N_0 \to \infty$. Assume also that we have an estimator $\hat{\delta}$ of $\delta$ such that $\max_{s \in \mathcal{I}_0 \cup \mathcal{I}_1} |h(Z_s,\hat{\delta})-h(Z_s, {\delta})| = o_{\mathbb{P}}(1)$ as $N_0 \to \infty$. Then, (i) $\widehat \alpha$ converges in probability to $\alpha + \widetilde{W}$, and (ii) if the null is true, then $\lim_{N_0,B\to \infty}\mathbb{E}[\phi_{\mbox{\tiny FP}} ] = \tau$. 

\end{proposition_no}

This approach is well-suited for settings in which $Y_{st}$ is the state $\times$ time aggregate of individual-level observations $Y_{ist}$. Let $Z_s$ be the number of individual-level observations in state $s$. FP consider the case in which heteroskedasticity arises only from  variation in the number of observations per unit.  In this case, we should expect $\mathbb{V}(W_{s})$ to be a decreasing function of $Z_s$. 
FP show that, under a wide range of structures on the within-unit correlations, $\mathbb{V}(W_{s} )$ would be given by $A + B/Z_s$, for parameters $A,B \geq 0$. Note that the bounding conditions on $h(Z_s,\delta)$ are satisfied in this setting if either $A > 0$, or $B>0$ and the sequence $(Z_s)_{s \in \mathcal{I}_0 \cup \mathcal{I}_1}$ is bounded uniformly as $N_0 \to \infty$. In this case, the idea is to estimate $A$ and $B$ using the residuals from the control units, and then re-scale the residuals to the control units in order to approximate the marginal distributions  of $W_s$ for the treated. Then we can use these distributions to compute critical values. This approach can be used when we have access to the individual-level data, or when we have only aggregate data (provided that we have access to information on the number of observations per unit). 

\begin{remark}
\label{Remark_AF}
    Since these methods rely on information from the control units to assess uncertainty, assuming that treatment effects are non-stochastic is crucial, because control units do not provide any information about stochastic treatment effect heterogeneity {(see \cite{alvarez2025inferencetreatedunits} for further discussion)}. Still, if treatment effects are stochastic, we can consider alternative interpretations for these methods, following \cite{AF2025}. First, these methods are valid  for testing sharp nulls such as $H_0: Pr\left(\frac{1}{N_1} \frac{1}{T-t^\ast} \sum_{s \in \mathcal{I}_1}\sum_{t \in \mathcal{T}_1} \alpha_{st} = \alpha_0 \right)=1$. Second, if  treatment effect heterogeneity is independent from errors $W_s$, then these methods are valid for inference on the realized treatment effects. That is, inference would be conditional not only on treatment assignment, but also on the realized values of $\alpha_{st}$. Finally, regardless of the dependence between treatment effect heterogeneity and errors, these methods can be inverted to construct prediction intervals (rather than confidence intervals). This logic also applies to the inference methods that will be discussed in Sections \ref{Sec_single_treated} and \ref{Sec: alternatives}, so we can interpret those methods in the same three ways when treatment effects are stochastic. 

\end{remark}

\section{Setting with spatial correlation}

We consider now the case in which errors may be spatially correlated. We consider the following assumptions.

\begin{assumption}{} \label{As_W}
\normalfont
 (i) $W_s = h(Z_s;\delta) \xi_s$, where $\xi_s$ is equally distributed for all  $s \in \mathcal{I}_1 \cup \mathcal{I}_0$, with a common distribution function $F_\xi$ that is absolutely continuous with finite first moment; (ii)  there exists a constant $\underline h > 0$ not depending on $N_0$ such that $\min_{s \in   \mathcal{I}_0 \cup \mathcal{I}_1} h(Z_s;\delta) \geq \underline h$ for all $N_0 \in \mathbb{N}$; (iii)  $\frac{1}{N_0} \sum_{s \in \mathcal{I}_0}W_s \buildrel p \over \rightarrow 0$, $\frac{1}{N_0} \sum_{s \in \mathcal{I}_0}|W_s| = O_{\mathbb{P}}(1)$  and $\frac{1}{N_0} \sum_{s \in \mathcal{I}_0}  \mathbbm{1}\{ \xi_s \leq c \}  \buildrel p \over \rightarrow  F_\xi(c)$ for every $c \in \mathbb{R}$ when $N_0 \rightarrow \infty$; and (iv) the estimator $\widehat \delta$ for $\delta$ is such that $\max_{s \in  \mathcal{I}_1 \cup \mathcal{I}_0} | h(Z_s;\widehat \delta) - h(Z_s; \delta)| = o_{\mathbb{P}}(1)$ when $N_0 \rightarrow \infty$. 
\end{assumption}

Assumptions \ref{As_W}(i) and \ref{As_W}(ii)  restrict the marginal distribution of the treated and control units, as in CT and FP. With these assumptions, the residuals of the control units become informative about the distribution of the errors of the treated units, which is the main insight from CT.  If we set $h(Z_s,\delta)$ constant, then these assumptions imply that the marginal distribution $W_s$ is the same for both treated and control units.

 Assumption \ref{As_W}(iii) is a high-level assumption that allows for spatially correlated shocks, but restricts such dependence so that we can apply a law of large numbers when we consider the control units. This will be satisfied, for example, if we assume strong mixing conditions in the cross section (see Theorem 3 from \cite{JENISH200986}). More generally, however, laws of large numbers are known to hold for a wide range of spatially dependent processes \citep{Jenish2012}.  For simplicity, we refer to this assumption in the text  as a weak dependence assumption. Importantly, Assumption \ref{As_W} allows for arbitrary spatial correlation among the treated units. Finally, Assumption \ref{As_W}(iv) states that we can consistently estimate the parameters of the heteroskedasticity.

Since we focus on settings in which researchers do not have information on what generates the spatial correlation or they are not willing to assume such structure, we do not need to model in detail the sources of spatial correlation. As a concrete  example, we can consider a setting in which the $N_1$ treated units are closely located geographically, but we have a larger number of control units in different locations. In this case, if spatial correlation goes to zero when geographical distance increases, then we would have  Assumption \ref{As_W}(iii) satisfied, even though we may have arbitrarily strong spatial correlation among the $N_1$ treated units. While it is natural to think about spatial correlation based on geographical distance, this may not be the case in relevant empirical applications. For example, we may have that units with similar industry shares  have more correlated errors. Importantly, we consider a setting in which the applied researcher may be unaware or may not have information on the relevant distance metrics in the cross section, which is common in DID applications  \citep{FermanJAE}.

Under Assumptions \ref{As_id} and  \ref{As_W}, it follows again that $\widehat \alpha$ is unbiased, and that, when $N_1$ is fixed and $N_0 \rightarrow \infty$,  $\widehat \alpha$ converges in probability to $\alpha + \widetilde{W}$. We now consider different approaches for testing the null $H_0: \alpha = \alpha_0$ {and for constructing confidence intervals}.

\subsection{Case with $N_1=1$: {CT and FP remain valid}} \label{Sec_single_treated}

When $N_1=1$,  the inference methods proposed by CT and FP can remain valid even if we allow for spatial correlation, and even when we do not have information on the relevant distance metric. The main intuition is that, under Assumption \ref{As_W}, the asymptotic distribution of $\widehat \alpha$ depends only on $W_1$, and the distribution of $W_1$ can still be asymptotically approximated using the residuals from the controls, so we can use that to construct critical values.

Let $\widehat F_\xi (c) = {N_0}^{-1} \sum_{s \in \mathcal{I}_0} \mathbbm{1} \{ \widehat W_s/h(Z_s,\widehat \delta ) \leq c \}$, where $\widehat \delta$ is an estimator for $\delta$. We first show that $\widehat F_\xi (c)$  approximates the distribution of $\xi_s$ if  $\widehat \delta$ is consistent.

\begin{proposition} \label{Prop_N1_FP}
 Suppose Assumptions \ref{As_id} and \ref{As_W} hold. Then, as $N_0 \rightarrow \infty$, $\widehat F_\xi (c)$ converges in probability to $F_\xi(c)$, uniformly over $c \in \mathbb{R}$. 
 \begin{proof}
     See Appendix \ref{Proof_N1}.
 \end{proof}
\end{proposition}

The proof of Proposition \ref{Prop_N1_FP} is similar to the proof of Proposition 2 from CT.   In this setting with $N_1=1$, the inference method proposed by FP would approximate the asymptotic distribution of $\widehat \alpha$ with the empirical distribution of  $\{ h(Z_1,\widehat \delta) \widehat W_s / h(Z_s,\widehat \delta) \}_{s \in \mathcal{I}_0}$,  to construct the critical values. It immediately follows from Proposition \ref{Prop_N1_FP} that the inference method proposed by FP remains valid for the case with $N_1=1$, even when we may have spatial correlation.

\begin{corollary} \label{Cor_N1}

Suppose we have data on $\{Y_{s1},...,Y_{sT},Z_s\}_{ \mathcal{I}_1 \cup \mathcal{I}_0}$, and that Assumptions \ref{As_id} and \ref{As_W} hold. Consider the case where $N_1=1$ and let $\phi_{\text{FP}}$ denote the decision rule of the test that rejects $\alpha=\alpha_0$ if $\hat{F}_\xi\left(\frac{\hat{\alpha}-\alpha_0}{h(Z_1;\hat{\delta
})}\right) <\tau/2$ or $\hat{F}_\xi\left(\frac{\hat{\alpha}-\alpha_0}{h(Z_1;\hat{\delta
)}}\right) > 1- \tau/2$ . Then, under the null,  $\lim_{N_0\to \infty} \mathbb{E}[\phi_{\mbox{\tiny FP}}] \rightarrow \tau$.
\begin{proof}
     See Appendix \ref{proof_Cor}.
\end{proof}
\end{corollary}

 Note that if we set $h(Z_s,\delta)$ constant, and $h(Z_s,\widehat \delta)=1$, then Proposition \ref{Prop_N1_FP} and Corollary \ref{Cor_N1} imply that the standard procedure proposed by CT in their running model is also asymptotically valid when $N_1=1$ and $N_0 \rightarrow \infty$, even when we have spatial correlation.

\subsection{Case with $N_1>1$: inference problems}  \label{Problems}

When $N_1>1$, spatial correlation can lead to relevant size distortions if we rely on the  methods proposed by CT and FP. For simplicity, consider the case in which $N_1=2$, and  $\{W_1,W_2\}$ is multivariate normally distributed with  correlation $\rho$. Consider also the case in which $h(Z_s,\delta)$ is constant. Under Assumption \ref{As_W}, $\widehat \alpha \buildrel p \over \rightarrow \alpha + \widetilde{W}$, where  $\widetilde{W} \sim N(0,2^{-1} (1 + \rho) \mathbb{V}(W_s))$. However, if $W_s$ is weakly dependent for the control units, when we consider two random draws from  $\{\widehat W_s \}_{s \in \mathcal{I}_0}$ to recover the distribution of $\widetilde{W}$, the correlation between these draws would converge to zero when $N_0 \rightarrow \infty$. As a consequence, the  approach proposed by CT would recover a distribution for $\widetilde{W}$ that is normal with a variance  $2^{-1} \mathbb{V}(W_s)$. In this case, critical values would be too small, leading to over-rejection. The same problem applies for the inference method proposed by FP.

\section{Case with $N_1>1$: {a menu of solutions} } \label{Sec: alternatives}

We consider different alternatives that allows for spatial correlation, even when the applied researcher does not have information on the relevant source of spatial correlation. All alternatives are based on the fact that, under Assumptions \ref{As_id} and \ref{As_W}, we can consistently estimate the marginal distribution of  $W_s$ for all treated units, and they differ in terms of the required assumptions for the dependence between the errors of the treated units. Table \ref{tab:methods} summarizes these alternatives, highlighting the main assumptions we need to impose on the spatial correlation among the treated units (in addition to Assumptions \ref{As_id} and \ref{As_W}).

\begin{table}[h!]
\centering
\caption{Menu of methods for inference with spatial correlation}
\begin{tabular}{lll}
\hline
 & \textbf{Main assumptions} & \textbf{Algorithm} \\
\hline
\textbf{Method 1:} & 
No restriction on the spatial correlation & 
 Algorithm \ref{alg_es}  \\ \\
\textbf{Method 2:} & 
Positive dependence (Assumption \ref{ass_psrd}) & 
\pbox{0.35\textwidth}{Conduct inference using the confidence interval in equation \ref{eq_refined}} \\ \\
\textbf{Method 3:} &  \pbox{0.5\textwidth}{Restrictions on the tails of $\widetilde{W}$ (Assumption \ref{As_reg})} & Algorithm \ref{conservative_1}  \\ \\
\textbf{Method 4:} & \pbox{0.5\textwidth}{Restrictions on the tails of $\widetilde{W}$ and on the within/between units spatial correlation (Assumption \ref{As_reg_cons2}); only valid when $Y_{st}$ is the aggregate of individual-level observations} & Algorithm \ref{conservative_2}\\
\\
\hline
\end{tabular}
\label{tab:methods}
\end{table}

\subsection{Method 1:  arbitrary dependence}
\label{bounds_arbitrary}
Our first approach seeks to bound the quantiles of $\widetilde{W}$, without imposing any assumption on the dependence structure between the errors of treated units. Indeed, notice that, if one is able to find an upper bound $\overline{c}_{1-\tau/2}$ (lower-bound $\underline{c}_{\tau/2}$) to the $1-\tau/2$ ($\tau/2$) quantile of  $\widetilde{W}$, then a test that rejects the null if $\hat{\alpha}-\alpha_0$ is above a consistent estimator of the upper bound (or below a consistent estimator of the lower bound) will have an asymptotically correct size. Since, for continuous distributions, a lower bound to the $u$-quantile  of $\widetilde{W}$ can be obtained by finding an upper bound to $Q_{-\widetilde{W}}(1-u)$, without loss of generality, we focus our discussion on the construction of an upper bound to $Q_{\widetilde{W}}(u)$ for a given $u \in (0,1)$.

We consider the combination of two simple upper bounds to $Q_{\widetilde{W}}(u)$ as a means of providing conservative inference when one does not wish to restrict the dependence structure across the $\xi_s$. The first upper bound is based on a union bound correction:\footnote{This bound follows from the observation that, for any sequence $(X_j)_{j=1}^k$ of random variables: $\mathbb{P}\left[\sum_{j=1}^k X_j > \sum_{j=1}^k Q_{X_j}\left(1- \frac{(1-u)}{k}\right)\right] \leq \sum_{j=1}^k \mathbb{P}\left[X_j> Q_{X_j}\left(1-\frac{(1-u)}{k}\right) \right] \leq 1-u$.}

$$Q_{\widetilde{W}}(u) \leq \sum_{s \in \mathcal{I}_1} \frac{1}{N_1}h(Z_s;\delta) Q_{\xi}\left(1 -  \frac{(1-u)}{N_1}\right) =: \overline{C}^{UB}_{\widetilde{W}}(u)\, . $$

Even though relying on a union bound correction ensures proper size control, it can lead to very conservative inference, especially as $N_1$ increases. We thus consider a second, dimension-independent, bound. This bound follows from the observation that, if $F_\xi$ is continuous, then:
$$Q_{\widetilde{W}}(u) \leq \mathbb{E}[\widetilde{W}| \widetilde{W} \geq Q_{\widetilde{W}}(u) ]\leq \sum_{s \in \mathcal{I}_1}\frac{1}{N_1}h(Z_s;\delta) \mathbb{E}[\xi|\xi \geq Q_\xi(u)] =: \overline{C}^{ES}_{\widetilde{W}}(u)\,,$$
where the second inequality follows from observing that $\mathbb{E}[\widetilde{W}| \widetilde{W}\geq  Q_{\widetilde{W}}(u) ]$ is an expected-shortfall (ES) type of  measure, and that this measure is sublinear or coherent \citep{Rockafellar2002}.\footnote{Specifically $\mathbb{E}[\widetilde{W}| \widetilde{W}\geq  Q_{\widetilde{W}}(u) ]$ corresponds to the $\operatorname{CVaR}_{u}^-(\widetilde{W})$ measure of \cite{Rockafellar2002}. This measure satisfies $\operatorname{CVaR}_{u}^-(\widetilde{W})\leq \operatorname{CVaR}_{u}(\widetilde{W})$, where $\operatorname{CVaR}_{u}(\widetilde{W})$ is the expected-shortfall or conditional Value-at-Risk measure that is sublinear \citep[Corollary 11]{Rockafellar2002}. Sublinearity implies that  $\operatorname{CVaR}_{u}(\widetilde{W}) \leq \sum_{i=1}^N\frac{1}{N_1}h(Z_i;\delta) \operatorname{CVaR}_{u}(\xi_i)$, and the expression in the main text then follows from observing that, for continuous distributions, the CVaR and CVaR$^-$ coincide. } The upper bound $\overline{C}_{ES}(u)$ coincides with the one obtained in Theorem 2.6 of \cite{Wang2013}, which is known to be pointwise-optimal under shape restrictions on $F_\xi$. Given that the bound is not, in general, sharp, though, we thus propose to combine it with the union bound correction, yielding the bound

$$Q_{\widetilde{W}}(u) \leq \overline{C}^{UB}_{\widetilde{W}}(u) \land \overline{C}^{ES}_{\widetilde{W}}(u)=: \overline{Q}_{\widetilde{W}}(u)\, .  $$

Notice that, given a consistent estimator of $F_\xi$, we can estimate all the elements needed to compute $\overline{Q}_{\widetilde{W}}(u)$. {This leads us to consider Algorithm \ref{alg_es}, where rejection of the null $H_0: \alpha = \alpha_0$ at significance level $\tau$ is based on plug-in estimates of the bounds to the $\tau/2$ and $(1-\tau/2)$ quantiles of $\widetilde{W}$.}

{\singlespacing
\renewcommand{\thealgocf}{M1}
\begin{algorithm}[h]
Run the DID estimator, and store $\widehat \alpha$ and the residuals  $\widehat W_s = \frac{1}{T-t^\ast} \sum_{t \in \mathcal{T}_1}\hat \eta_{st} -  \frac{1}{t^\ast} \sum_{t \in \mathcal{T}_0} \hat \eta_{st}$ for all $s \in \mathcal{I}_0$ \;
 Estimate $\delta$ using the residuals from the controls, and compute the normalized residuals $\widehat \xi_s = \widehat W_s / h(Z_s,\widehat \delta)$, $s \in \mathcal{I}_0$ \;
 Compute the empirical distribution of the normalized residuals $\hat{F}_\xi$ and the associated empirical quantile functions $\hat{Q}_\xi$ \;
Estimate the bounds:
$$ \hat U = \sum_{s\in \mathcal{I}_1} \frac{1}{N_1}h(Z_s;\hat \delta)  \times \min\left\{\hat{Q}_\xi\left(1-\frac{\tau}{2N_1}\right), \frac{1}{\tau/2}\int_{\hat Q_\xi\left(1-\frac{\tau}{2}\right)}^\infty v \hat F_\xi(dv) \right\}$$
$$ \hat L =  \sum_{s \in \mathcal{I}_1} \frac{1}{N_1}h(Z_s;\hat \delta)  \times \max\left\{\hat{Q}_\xi\left(\frac{\tau}{2N_1}\right), \frac{1}{\tau/2}\int_{-\infty}^{\hat Q_\xi\left(\frac{\tau}{2}\right)}v \hat F_\xi(dv) \right\}$$
 Reject the null if $\hat \alpha - \alpha_0> \hat U$ or $\hat \alpha - \alpha_0< \hat L$.
	\caption{Test based on the union bound and ES bounds}
     \label{alg_es}
\end{algorithm} }

The following proposition establishes asymptotic conservativeness of Algorithm \ref{alg_es}.  Importantly, we do not impose impose any assumption on the spatial dependence of the errors of treated units.

\begin{proposition}
\label{prop_es}
    Suppose that that Assumptions \ref{As_id} and \ref{As_W} hold. Let ${\phi}_{1}$ denote the decision rule from Algorithm \ref{alg_es}. Then, under the null, as $N_0 \to \infty$, $\limsup_{N_0\to \infty}\mathbb{E}[\phi_{1}]\leq \tau $.
    \begin{proof}
        See Appendix \ref{proof_prop_es}.
    \end{proof}
\end{proposition}

\begin{remark}[Sharper bounds when $N_1=2$] When there are two treated units, sharper bounds to $Q_{\widetilde{W}}(u)$ than $\bar{Q}_{\widetilde{W}}(u)$ can be recovered by relying on a construction due to \cite{Makarov1982}. Appendix \ref{sec_makarov} establishes the asymptotic conservativeness of an inference procedure based on a plug-in version of such \cite{Makarov1982} bound. We also provide sufficient conditions for this bound to yield the same conclusions as the union bound correction.
\end{remark}

\begin{remark}[Alternative upper bounds when $N_1 > 2$]
 For the general case where $N_1>2$, upper bounds to $Q_{\widetilde{W}}(u)$ have also been obtained by \cite{Puccetti2013} and \cite{Embrechets2013}. These bounds have been shown to be pointwise-optimal in the homoskedastic case, under some assumptions. However, their computation can be cumbersome, and optimality in the heteroskedastic case is yet to bet established. For tractability, we thus opt to construct bounds by combining the union bound and expected-shortfall corrections, the latter of which is known to collapse to the correction of \cite{Puccetti2013} and \cite{Embrechets2013} in special cases.
\end{remark}

\subsection{Method 2:  positive dependence of the errors}
\label{Bounds_positive}
The bounds in the previous sections do not impose any assumption on the dependence structure of the $\xi_s$. In this section, we show how we may shrink these bounds under a mild positive dependence assumption on the $\xi_s$. Following \cite{Benjamini2001}, we say a set $D \subseteq \mathbb{R}^{N_1}_+$ is increasing if $x \in D$ and $y \geq x$ imply $y \in D$. We then make the following assumption on the joint distribution of the $\xi_s$.

\begin{assumption}
\label{ass_psrd}
    The joint distribution of $(|\xi_1|,\ldots, |\xi_{N_1}|)$ exhibits positive dependence, in the sense that, for any $s\in \mathcal{I}_1$ and $D \subset \mathbb{R}^{N_1}_+$ increasing, the map $x \mapsto \mathbb{P}[(|\xi_1|,\ldots, |\xi_{N_1}|) \in {D}||\xi_x|=x]$ is nondecreasing. 
\end{assumption}

Intuitively, Assumption \ref{ass_psrd} implies that extremer values of one of the $\xi_s$ weakly predict more extreme values of the remainder $\xi_{s'}$, $s'\neq s$.

 Notice that, in light of Corollary \ref{Cor_N1}, asymptotic inference on the individual treatment effects ${\alpha}_s = \frac{1}{T_1}\sum_{t\in \mathcal{T}_1}\alpha_{s,t}$ can be performed separately for each treated unit $s \in \mathcal{I}_1$ by computing a DID estimator that leverages data from the control group and a single treated unit $i$. Under Assumption \ref{ass_psrd}, we can then construct a $(1-\tau)$ confidence set for the vector of individual treatment effects $( \alpha_s)_{s\in \mathcal{I}_1}$, by inverting the \cite{Benjamini1995} procedure for control of the False Discovery Rate (FDR). Specifically, one constructs this set by collecting each vector $(a_s)_{s \in \mathcal{I}_1} \in \mathbb{R}^{N_1}$ such that the procedure does not reject \textbf{any} null when testing the nulls $\alpha_s = a_s$, $s \in \mathcal{I}_1$, at significance level $\tau$. Given that, under Assumption \ref{ass_psrd} and when all the nulls are true, the probability of rejecting at least one null is no greater than $\tau$, this inversion procedure yields a valid joint confidence region for the vector of individual effects.  One can then construct a confidence set for the average effect $\alpha = \frac{1}{N_1}\sum_{s \in \mathcal{I}_1}\alpha_s$ by the method of \emph{projection}: we  calculate and store $\sum_{s \in \mathcal{I}_1} \frac{1}{N_1} a_s$ for each $(a_s)_{s \in \mathcal{I}_1}$ in the confidence set for individual effects \citep{Scheffe1958,Dufour1990,gafarov2016projection,Freyberger2018}. 

While we leave the steps of this construction to Appendix \ref{Appendix_MHT}, we observe that the resulting confidence interval for $\alpha$ is given by:

\begin{equation}
    \label{eq_bh}
\mathcal{I}_{BH} = \left[\hat{\alpha}-c^*, \hat{\alpha}+c^*\right]\, ,
\end{equation}
where $c^* = \sum_{j=1}^{N_1}\frac{1}{N_1}h(Z_{(N_1+1-j)};\hat{\delta}) \hat{Q}_{|\xi|}\left(1-j\frac{ \tau}{N_1}\right)$, with $h(Z_{(j)};\hat \delta)$ denoting the $j$-th smallest element of the set $\{h(Z_s;\hat \delta): s \in \mathcal{I}_1\}$. Here, $\hat{Q}_{|\xi|}$ denotes the empirical quantile function of $|\xi|$.

The length of confidence interval given  by \eqref{eq_bh} can be smaller than the confidence interval obtained from inverting Algorithm \ref{alg_es}, especially if $N_1$ is small. However, as $N_1$ increases, the dependence of the limits on $N_1$ can make confidence intervals based on inversion of Algorithm \ref{alg_es} more appealing. Since the latter are always valid, we thus propose the refined interval:

\begin{equation}
    \label{eq_refined}
\mathcal{I}_{\text{Refined}} =  \begin{cases}
\mathcal{I}_{BH} & \text{if } 2c^* < \hat U - \hat L
   \\ 
       [\hat{\alpha}-\hat{U}, \hat{\alpha}-\hat{L}] & \text{if } 2c^* \geq \hat U - \hat L
  \tag{M2}
\end{cases}\, ,
\end{equation}
where $(\hat{U},\hat{L})$ denotes the critical values of Algorithm \ref{alg_es}.\footnote{If $N_1=2$, another option is to use the critical values from the Makarov bounds of Algorithm \ref{alg_makarov} in Appendix \ref{sec_makarov} instead of those of Method \ref{alg_es}.}

The following proposition summarizes the properties of the refined interval. 

\begin{proposition}
\label{prop_mht}
    Suppose that Assumptions \ref{As_id}, \ref{As_W} and \ref{ass_psrd} hold. Then $\liminf_{N_0 \to \infty}\mathbb{P}[\alpha \in \mathcal{I}_{\text{Refined}}]\geq 1-\tau$.

    \begin{proof}
        See Appendix \ref{proof_mht}.
    \end{proof}
\end{proposition}

Given the way we construct the confidence interval in Method 2, we can guarantee that its length will be weakly smaller relative to that of Method 1. The simulations in Section \ref{simulations} show that the confidence intervals from Method 2 may be strictly smaller than those from Method 1.

\subsection{Method 3: restrictions on the tails of $\widetilde{W}$}
\label{Sec_cons1}
As a third alternative, we show that under intuitive and mild assumptions on the tails of $\widetilde{W}$ it is possible to construct more informative bounds relative to Methods 1 and 2. We calculate bounds for the quantiles of $\widetilde{W}$ considering that the worst-case scenario for spatial dependence of the $(W_1,...,W_{N_1})$ is given by the comonotone copula. This means calculating critical values based on the empirical distribution of  $$\widehat H(c) = \frac{1}{N_0} \sum_{s' \in \mathcal{I}_0}  \mathbbm{1}  \left\{ N_1^{-1} \sum_{s \in \mathcal{I}_1} h(Z_s,\widehat \delta) \widehat \xi_{s'} < c  \right\}.$$ 

In the particular case in which $h(Z_s,\delta)$ is constant,  we calculate critical values based on the marginal distribution of $W_s$. Since $\mathbb{V}(\widetilde{W}) \leq \mathbb{V}({W_s})$, we know that our critical values are based on a distribution that has a variance no higher than the asymptotic variance of $\hat \alpha$. However, this is not sufficient to guarantee that the test is asymptotically valid. We therefore, consider a high-level assumption that guarantees that this proposed method is valid, and then we discuss the plausibility and sufficient conditions for this assumption.

\begin{assumption} \label{As_reg}
\normalfont

$ \mathbb{P}\left( \{ \widetilde W  > c_{1-\tau/2}  \} \cup  \{ \widetilde W  < c_{\tau/2}  \} \right) \leq \tau $, where  $c_{u}$ is the $u$-quantile of the distribution of $N_1^{-1} \sum_{s \in \mathcal{I}_1} h(Z_s, \delta)  \xi$.

\end{assumption}

Consider the simpler case in which $h(Z_s,\delta)$ is constant. Then this regularity condition simply means that, regardless of the spatial correlation among the treated units, the probability of having extreme values for the average of the treated units, $\widetilde W$, is weakly smaller than the probability of having extreme values for a single draw of $W_s$. Note that Assumption \ref{As_reg} is satisfied if $\{W_s\}_{s \in \mathcal{I}_1}$ is multivariate normal. In this case, $\widetilde W$ would also be normally distributed, and we have that   $\mathbb{V}(\widetilde W) \leq \mathbb{V}(W_s)$ irrespectively of the spatial correlation among the treated units. Therefore,  $\widetilde W$ will be less likely to attain extreme values than $W_s$.   The same intuition remains valid if  $h(Z_s,\delta)$ is not constant.

In Appendix \ref{Appendix: W tilde}, we provide further evidence on the plausibility of Assumption \ref{As_reg} beyond the case of multivariate normal distributions. Appendix \ref{app_joint_list} surveys several results available in the literature showing that Assumption \ref{As_reg} also holds for a wide range of joint distributions of $(\xi_s)_{s\in \mathcal{I}_1}$. These include elliptical joint distributions and, under some restrictions on significance values and the class of \textit{copulae} that specify the dependence structure between the $(\xi_s)_{s\in \mathcal{I}_1}$, also settings where the marginal distributions of $\xi_s$ feature heavy-tails. Appendix \ref{App_copulas} provides an approach to assess the validity of Assumption \ref{As_reg} in particular applications under the assumption that the true dependence structure belongs to the family of Gaussian \textit{copulae}. We find evidence that this assumption holds in the  ACS data on which we base our MC simulations in Section \ref{simulations} and in our empirical illustration in Section \ref{illustration}. Moreover, in case we find Gaussian \textit{copulae} in which Assumption \ref{As_reg} is not satisfied, we show how to adjust the test so that it is valid under the assumption that the dependence structure follows a Gaussian copula. Appendix \ref{Appendix_simple_check} presents an alternative empirical exercise to assess the plausibility of this assumption in the ACS data, based on an estimated model for the spatial correlation. Finally, to provide further intuition on Assumption \ref{As_reg},   Appendix \ref{Appendix_reg} presents an example in which this assumption may not hold. To engineer such example,  we consider a distribution for $W_s$ that is bimodal, and such that the two peaks are very far apart, which is not something we should expect in common empirical applications. Importantly, we note that applied researchers can evaluate the marginal distribution of $\widehat W_s$ in their empirical applications. Therefore, if the marginal distribution does not exhibit such large peaks  in the tails of the distribution, then they should have more confidence that Assumption \ref{As_reg} is valid. 
    
Overall, we see Assumption \ref{As_reg} as intuitive and plausible in most empirical applications. Given Assumption \ref{As_reg}, the proposed method can be implemented using Algorithm \ref{conservative_1}.

{\singlespacing
\renewcommand{\thealgocf}{M3}
\begin{algorithm}[h]
Run the DID estimator, and store $\widehat \alpha$ and the residuals  $\widehat W_s = \frac{1}{T-t^\ast} \sum_{t \in \mathcal{T}_1}\hat \eta_{st} -  \frac{1}{t^\ast} \sum_{t \in \mathcal{T}_0} \hat \eta_{st}$ for all $s \in \mathcal{I}_0$ \;
 Estimate $\delta$ using the residuals from the controls, and compute the normalized residuals $\widehat \xi_s = \widehat W_s / h(Z_s,\widehat \delta)$, $s \in \mathcal{I}_0$ \;
  Compute the empirical distribution of the normalized residuals $\hat{F}_\xi$ and the associated empirical quantile function $\hat{Q}_{\xi}$ \;
 Reject the null if $\hat{\alpha}-\alpha_0>\sum_{s\in \mathcal{I}_1}\frac{1}{N_1}h(Z_s;\hat \delta)\hat{Q}_{\xi}(1-\tau/2)$ or $\hat{\alpha}-\alpha_0<\sum_{s\in \mathcal{I}_1}\frac{1}{N_1}h(Z_s;\hat \delta)\hat{Q}_{\xi}(\tau/2)$
	\caption{Method 3 (restrictions on the tails of $\widetilde{W}$)}
    \label{conservative_1}
\end{algorithm}
}

It follows directly from Proposition \ref{Prop_N1_FP} and Assumption \ref{As_reg} that this modified test asymptotically controls for size under these assumptions.  

\begin{proposition} \label{Prop_conservative_FP}

Suppose we have data on $\{Y_{s1},...,Y_{sT},Z_s\}_{ \mathcal{I}_1 \cup \mathcal{I}_0}$, and that Assumptions \ref{As_id}, \ref{As_W}, and \ref{As_reg} hold. Let $\phi_{3}$ be an indicator variable equal to one if we reject the null in Algorithm \ref{conservative_1}. Then, if the null is true,  $\limsup_{N_0\rightarrow \infty} \mathbb{E}[\phi_{3}  ] \leq \tau$.

\end{proposition}
\begin{proof}
    Proposition \ref{Prop_N1_FP} and Lemma \ref{lemma_uniform_qt} in the Appendix imply that, as $N_0 \to \infty$: (i) $\sum_{s\in \mathcal{I}_1}\frac{1}{N_1}h(Z_s;\hat \delta)\hat{Q}_{\xi}(1-\tau/2) \overset{p}{\to} c_{1-\tau/2}$; and (ii) $\sum_{s\in \mathcal{I}_1}\frac{1}{N_1}h(Z_s;\hat \delta)\hat{Q}_{\xi}(\tau/2) \overset{p}{\to} c_{\tau/2}$. Consequently, it follows from the continuous mapping theorem that, under the null: $\phi_3 \overset{p}{\to} \mathbf{1}_{\{\widetilde{W} > c_{1-\tau/2}\}\cup \{\widetilde{W} < c_{1\tau/2}\}}$. It then follows from the bounded convergence theorem and Assumption \ref{As_reg} that, under the null: $\lim_{N_0 \to \infty}\mathbb{E}[\phi_3]= \mathbb{P}[\{\widetilde{W} > c_{1-\tau/2}\}\cup \{\widetilde{W} < c_{1\tau/2}\}]\leq \tau$.
\end{proof}

 In Appendix \ref{app_rank}, we show that confidence intervals obtained from Method 3 always result in smaller length than those obtained from Method 1. Moreover, if the distribution of the residuals $\hat \xi_s$ is symmetric, then we also have that confidence intervals obtained from the inversion of Method 3 are contained in those obtained from Method 2.  Finally, our simulation results from Section \ref{simulations} illustrate how Assumption \ref{As_reg}, which we see as intuitive and plausible, can go a long way in reducing the length of the confidence intervals in this setting.

\subsection{Method 4: bounding the across-unit correlations}\label{Sec_cons2}

For settings in which $Y_{st}$ represents averages of $M_{st}$ individual-level observations $Y_{ist}$, we show that it is possible to construct an alternative test that will generally be less conservative than Method 3.  In this case,  we assume that individual-level observations within the same unit are weakly more spatially correlated than individual-level observations in different units. Then, we can use information on the within-unit spatial correlation to bound the across-units spatial correlation. We can consider  either the case in which the econometrician  observes only unit aggregates $Y_{st}$ (but has information on $M_{st}$) or the case in which individual-level data $Y_{ist}$ is observed. For simplicity, consider the case in which $M_{st} = M_s$ for all $t$ {(see Appendix \ref{Appendix_cons2} for  the case in which $M_{st}$ varies across $t$).} We treat the sequence $\{M_{s}\}_{s \in \mathcal{I}_0 \cup \mathcal{I}_1}$ as fixed. 

In this setting, FP show that, under a wide range of structures on the within-unit correlations for the individual-level errors, we have that $\mathbb{V}(W_s ) = A + B/M_s$ for constants $A,B \geq 0$. Importantly, these parameters are informative about the within-unit correlations, and can be used to bound the across-unit correlations, under the assumption that within-unit correlations are stronger than the across unit ones.

Let $M_{\mbox{\tiny T}} = \sum_{s \in \mathcal{I}_1} M_s$ be the total number of individual-level observations in the treated units, and $M_{\mbox{\tiny C}} = \sum_{s \in \mathcal{I}_0} M_s$ the number of individual observations among the control units. We consider in this case the DID estimator weighted by $M_s$, $\widetilde \alpha$, so that 
\begin{eqnarray} \label{}
\widetilde \alpha = \alpha' + \frac{1}{M_{\mbox{\tiny T}} } \sum_{s \in \mathcal{I}_1} M_s W_s - \frac{1}{M_{\mbox{\tiny C}} } \sum_{s \in \mathcal{I}_0} M_s W_s ,
\end{eqnarray}
where $\alpha' \equiv \frac{1}{M_{\mbox{\tiny T}}}    \sum_{s \in \mathcal{I}_1}  M_s \left(\frac{1}{T-t^\ast}\sum_{t \in \mathcal{T}_1} \alpha_{st}\right)$ is the weighted average treatment effect across units and treated periods.  The weighted DID estimator in this case is numerically the same as the DID estimator using the individual-level data. 

We consider a version of Assumption \ref{As_W} for this specific setting. In particular, we consider that the observed covariate driving the heteroskedasticity is $M_s$ (the number of individual-level observations), and that $\mathbb{V}(W_s ) = A + B/M_s$.

\begin{assumption}{} \label{As_W_M}
\normalfont
(i) $W_s = \left(A + B/M_s\right)^{1/2} \xi_s$ for nonnegative constants $A,B$, where $\xi_s$ is equally distributed for all  $s \in \mathcal{I}_1 \cup \mathcal{I}_0$, with a distribution $F_\xi$ that is absolutely continuous with finite first moment; there exists $\underline{h}>0$, not depending on $N_0$, such that $\min_{j \in \mathcal{I}_0 \cup \mathcal{I}_1} \left\{A + B/M_j \right\} \geq \underline{h}$, for all $N_0 \in \mathbb{N}$; (iii) $\frac{1}{\sum_{s \in \mathcal{I}_0}M_s}   \sum_{s \in \mathcal{I}_0}M_s W_s \buildrel p \over \rightarrow 0$, $\frac{1}{M_{\tiny C}}\sum_{s \in \mathcal{I}_0}M_s |W_s| = O_{\mathbb{P}}(1)$ and $\frac{1}{N_0} \sum_{s \in \mathcal{I}_0}  \mathbbm{1}\{ \xi_s \leq c \}  \buildrel p \over \rightarrow  F_\xi(c)$ for every $c \in \mathbb{R}$; (iv) the (nonnegative) least squares estimators of $\widehat W_s^2$ on a constant and $1/M_s$ using only the control units are consistent for $A$ and $B$ as $N_0 \to \infty$.

\end{assumption}

Given Assumption \ref{As_W_M}, we have that $\widetilde \alpha \buildrel p \over \rightarrow \alpha' + \mathcal W $, where $\mathcal W = \frac{1}{M_{\mbox{\tiny T} }} \sum_{s \in \mathcal{I}_1} M_s W_s$. We propose the following idea for inference: we consider an aggregate treated unit that is the weighted average of the treated units. Then we run FP inference method using this aggregate treated unit and the controls, considering that it has $M_{\mbox{\tiny T}}$ observations.  Algorithm \ref{conservative_2} formalizes this idea.

{\singlespacing 
\renewcommand{\thealgocf}{M4}
\begin{algorithm}[h]
Run the weighted DID estimator, and store $\tilde \alpha$ and the residuals  $\widehat W_s = \frac{1}{T-t^\ast} \sum_{t \in \mathcal{T}_1}\hat \eta_{st} -  \frac{1}{t^\ast} \sum_{t \in \mathcal{T}_0} \hat \eta_{st}$ for all $s \in \mathcal{I}_0$ \;
Run a (nonnegative) least squares regression of $\widehat W_s^2$ on  a constant and $1/M_s$, to estimate $\widehat A$ and $\widehat B$ \;
Compute the normalized residuals $\widehat \xi_s = \widehat W_s / \sqrt{\widehat A + \widehat B/M_s}$, for $s \in \mathcal{I}_0$\;
Compute the empirical distribution of the normalized residuals $\hat{F}_\xi$ and the associated empirical quantile functions of $\xi$, $\hat{Q}_{\xi}$ \;

Reject the null if $ \widetilde \alpha - \alpha_0  > (\hat A + \hat{B}/M_T)^{1/2} \hat{Q}_{\xi}(1-\tau/2)$ or $  \widetilde \alpha - \alpha_0  < (\hat A + \hat{B}/M_T)^{1/2} \hat{Q}_{\xi}(\tau/2)$.
	\caption{Method 4 (bounding the across-unit correlations)}
    \label{conservative_2}
\end{algorithm}
}

To ensure the conservativeness of Algorithm \ref{conservative_2}, we consider the following high-level assumption stating that the probability that $\mathcal{W}$ attains extreme values is smaller than the probability that a hypothetical treated unit with $M_T$ observations would attain such values.

\begin{assumption} \label{As_reg_cons2}
\normalfont
$\mathbb{P}\left( \{ \mathcal W   > c_{1 -\tau/2}^* \} \cup   \{ \mathcal W   < c_{\tau/2}^* \} \right) \leq \tau $, where  $c_{u}^*$ is the $u$-quantile of the distribution of $[A + B/M_{\mbox{\tiny T} } ]^{1/2} \xi$.

\end{assumption}

 Suppose individual-level observations within the same unit are weakly more spatially correlated than individual-level observations in different units. In this case, we would have $\mathbb{V}(\mathcal{W}) \leq A + B / {M_{\mbox{\tiny T} }}$, which would guarantee that Assumption \ref{As_reg_cons2} holds if errors are multivariate normal. Appendix \ref{app_joint_list} shows that $\mathbb{V}(\mathcal{W}) \leq A + B / {M_{\mbox{\tiny T} }}$ also implies Assumption \ref{As_reg_cons2} in the more general case where $(W_s)_{s \in \mathcal{I}_1}$ follows an elliptical distribution; whereas Appendix \ref{App_copulas} shows how to assess the validity of this assumption in a particular application under Gaussian \textit{copulae}. We also present in Appendix \ref{app_micro}  an explicit construction that generates Assumption \ref{As_reg_cons2} from primitive conditions on individual-level data.  Overall, we see that as a reasonable and intuitive assumption in most settings, if we expect individual-level observations to be more correlated when they belong to the same unit relative to when they belong to different units.

Let $\phi_4$ be an indicator variable equal to one if we reject the null in Algorithm \ref{conservative_2}. The following result is an immediate consequence of Proposition \ref{Prop_N1_FP} and Assumption \ref{As_reg_cons2}. {The proof is analogous to that of Proposition \ref{Prop_conservative_FP} and, for the sake of brevity, we thus omit it.}

\begin{proposition} \label{Prop_conservative_2}
Suppose we have data on $\{Y_{s1},...,Y_{sT},M_s\}_{ \mathcal{I}_1 \cup \mathcal{I}_0}$, and that Assumptions \ref{As_id}, \ref{As_W_M} and \ref{As_reg_cons2} hold. Let $\phi_{4}$ be an indicator variable equal to one if we reject the null in Algorithm \ref{conservative_2}. Then, if the null is true,  $\limsup_{N_0 \to \infty} \mathbb{E}[\phi_4  ] \leq \tau$.

\end{proposition}

 In Appendix \ref{app_rank}, we show that the length of confidence intervals obtained from inversion of Method 4 is no greater than those obtained from a weighted version of Method 3 that targets $\alpha'$. Also, our simulation results from Section \ref{simulations} illustrate how Assumption \ref{As_reg_cons2}, which we again see as intuitive and plausible, can go a long way in reducing the length of the confidence intervals in this setting.

\subsection{Summary}

Overall, we provide a menu of alternatives that are valid for inference in this setting, even when spatial correlation is based on an unknown distance metric. These alternatives vary in the assumptions they rely on --- particularly in terms of the spatial correlation --- and in the length of the confidence intervals they generate. More specifically, we start with confidence intervals that do not impose any assumption on the spatial correlation among treated units, and show how mild and plausible assumptions may be used to reduce the confidence intervals' lengths. We summarize the relationship between the confidence intervals obtained from inversion of each inference method in Appendix \ref{app_rank}.  Our results highlight how being more specific about the worst-case scenario for spatial correlation -- as Methods 3 and 4 are -- can result in confidence interval length reductions, whilst still credibly preserving asymptotic coverage in the presence of unknown patterns of spatial correlation.

\begin{remark}

\label{Remark_distance}

If a distance metric is available and the researcher is willing to assume such distance metric is the relevant one for the spatial correlation, then other  available alternatives might present better power (for example, the inference method proposed in the Appendix of CT).\footnote{\cite{Conley1999} is another alternative usually considered in settings with spatial correlation with a known distance metric. However, this method does not perform well in settings with few treated units, for the same reason that cluster robust standard errors do not perform well with few treated clusters (see Section 2 of \cite{alvarez2025inferencetreatedunits}).} 
\end{remark}

\begin{remark}

\label{Remark_largeT}
Related to Remark \ref{Remark_distance}, another alternative to provide a more powerful test may be to  infer about the spatial correlation using the time series. For example, \cite{VOGELSANG2012303} and \cite{CWZ}.  FP also propose an alternative inference method in their section IV (not the one we reviewed in Section \ref{Sec_independent}) that allows for spatial correlation. However, such alternatives would require a large time series, while the alternatives we propose remain valid even when we have only one pre- and one post-treatment period. Given the survey from \cite{Roth}, settings in which the time series dimension is short are prevalent in DID applications.    

\end{remark}

\begin{remark}
\label{Remark_spatial}

The assumption  that $W_s$ is weakly dependent would not be satisfied if there are unobserved shocks affecting a non-negligible fraction of the controls (so that $\frac{1}{N_0} \sum_{s \in \mathcal{I}_0}W_s$ converges in probability to a non-degenerate random variable).  In this case, the DID residuals $\widehat W_s$ would not capture these  shocks, and $\widehat F_\xi(c)$ would underestimate the dispersion of the marginal distribution of $\xi_s$.  As a consequence, even the conservative tests from Section \ref{Sec: alternatives} may over-reject. Importantly, however,  the over-rejection  in this case would be no larger than the over-rejection for CT or FP.

\end{remark}

\section{Simulations with Real Datasets} \label{simulations}

We analyze the spatial correlation problem, and the proposed  conservative tests, in simulations based on the  ACS \citep{ipums}, at the Public Use Microdata Area (PUMA). We estimate a model for the spatial correlation in which the covariance between two PUMAs may depend  on whether they belong to the same state, and on the similarity between their industry compositions. We also allow for heteroskedasticity depending on population sizes (details in Appendix \ref{Appendix_MC}). This way, we can analyze three scenarios: (i) when the applied researcher ignores all spatial correlation; (ii) when he/she considers spatial correlation arising only from  geographical distance; and (iii) when he/she correctly considers that spatial correlation may arise from both geographical distance and industry composition. 

If we consider all pairs of PUMAs, the correlation between their errors in this estimated model is greater (in absolute value) than $0.05$ in only $2.5\%$ of the cases. Therefore, the weak dependence condition we consider seems reasonable in this setting. Still, we have PUMAs with spatial correlation as strong as $0.125$. Therefore, we may have a setting in which the $N_1$ treated PUMAs exhibit relevant spatial correlation. 

In line with our theoretical model, we consider  a setting in which we fix $N_1 \in \{1,2,\ldots,9,10\}$ PUMAs as the treated ones, and generate multivariate normal draws of $W_s$ with this estimated spatial dependence. To illustrate issues related to spatial correlation, we choose the PUMA with the strongest spatial correlation with some other PUMA in our dataset to be treated, and then iteratively assign treatment to $N_1-1$ PUMAs in the same state that are most similar in industry composition to previously selected PUMAs. Treatment effects are assumed to be zero. We consider seven different inference procedures: (i) the naive version of FP approach (which assumes errors are independent across PUMAs); (ii)  a parametric bootstrap that correctly specifies and estimates the heteroskedasticity structure and both sources of spatial correlation (this procedure is similar to the parametric bootstrap suggested in the Appendix of CT); (iii) a misspecified parametric bootstrap that estimates the heteroskedasticity structure due to group sizes and spatial correlation due to being in the same state, but ignores spatial correlation due to industry similarity; and  the conservative test presented in (iv) Section \ref{bounds_arbitrary} (Method 1), (v) Section \ref{Bounds_positive} (Method 2), Section \ref{Sec_cons1} (Method 3); and (vi) Section \ref{Sec_cons2} (Method 4). 

\begin{figure}[h]
			\caption{{\bf MC simulations}}  \label{Figure}

	\begin{subfigure}[b]{0.5\textwidth}
    		\caption*{A: Test size}
		\centering
		\includegraphics[width=\textwidth]{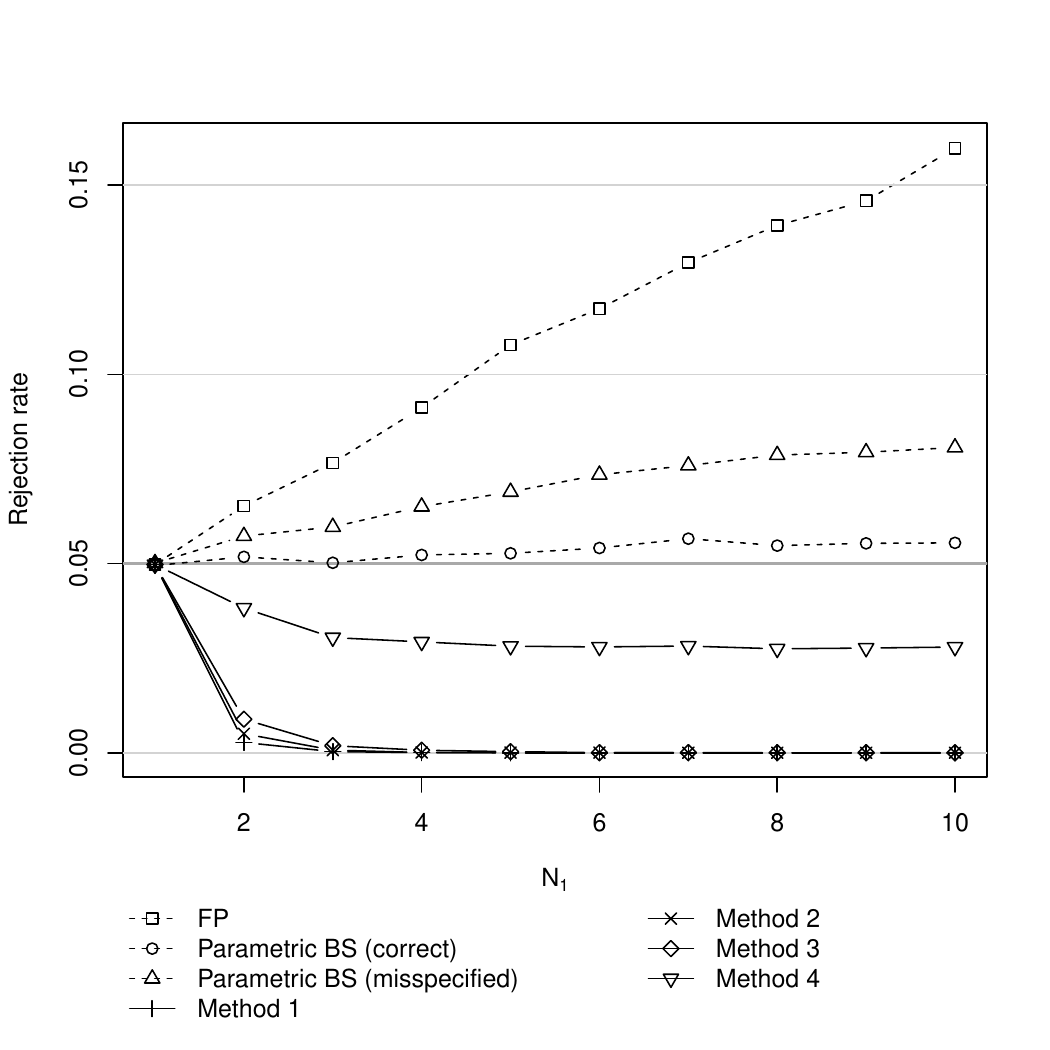}
	\end{subfigure}
				\begin{subfigure}[b]{0.5\textwidth}
		\centering
        		\caption*{B: Length of CI / length of unfeasible CI }

		\includegraphics[width=\textwidth]{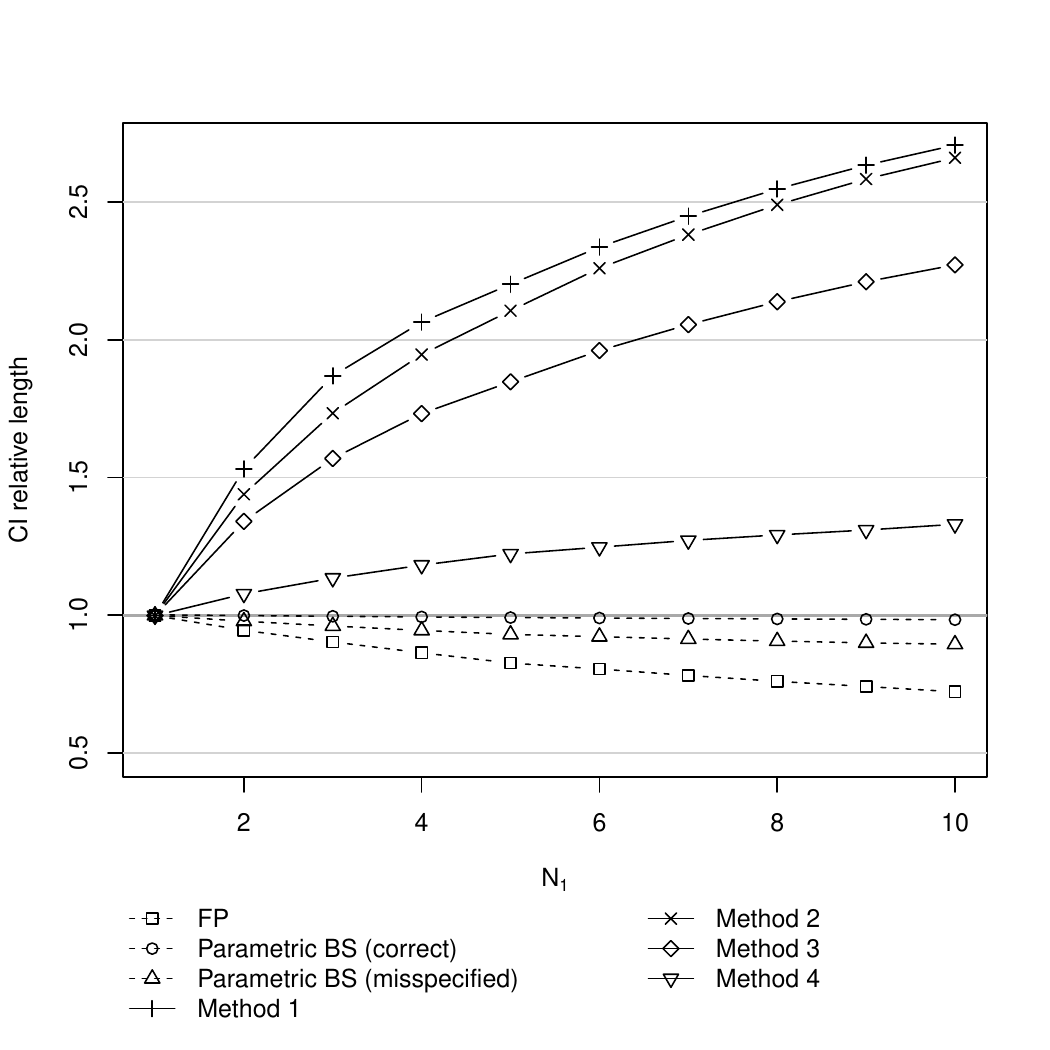}
	\end{subfigure}		

     \singlespacing\footnotesize
	\noindent \textit{Notes:} Panel A presents rejection rates for the seven different inference methods we discussed in Section \ref{simulations}, for different values of $N_1$. Tests are conducted nominally at the 5\% level.  Panel B presents information on the ratios between the average length  of nominal 95\% CIs obtained by test inversion relative to the length of the CI which uses the true sampling variance of $\widehat \alpha$.

\end{figure}

Figure \ref{Figure}.A presents rejection rates, as a function of $N_1$, of tests of the null of no average effect conducted nominally at the 5\% significance level, while Figure \ref{Figure}.B presents the ratio between the length of nominal 95\% confidence intervals obtained by inverting the corresponding test procedures and the length of a 95\% unfeasible confidence interval which uses the true sampling variance of $\widehat \alpha$ in the simulations.

Overall, these simulations highlight the main messages of the paper:  (1) the inference methods proposed by CT and FP remain valid when we have a single treated unit, even when errors are spatially correlated; (2) with $N_1>1$, these inference methods over-reject when errors are spatially correlated, and the over-rejection is increasing with $N_1$; (3) only accounting for the known sources of spatial correlation, e.g. only controlling for state correlation as the misspecified parametric bootstrap does, may not be sufficient to control size;  (4) the proposed conservative tests control for size, although they may be conservative when $N_1>1$; (5) the length of confidence intervals from Methods 1 and 2, which are not specific about the worst-case dependence structure, can be rather large; (6) Method 4 is able to provide less conservative tests by exploiting the structure of common empirical applications in which units are aggregates of individual-level observations to construct a more powerful test (even if we do not have information on the individual-level observations).

We note that, when $N_1 = 5$, the length of the confidence interval obtained by inverting Method 4 is only 22\% larger than the length of the unfeasible confidence interval. In contrast, Method 3 has an 85\% larger confidence interval than the unfeasible procedure. When $N_1=10$, Method 4 has  33\% larger confidence intervals, whereas confidence intervals with  Method 3 are more than 120\% larger. Methods 1 and 2 are even more conservative, with confidence intervals respectively 120\% and 111\% (171\% and 166\%) larger  when $N_1=5$ ($N_1=10$). Overall, these numbers illustrate the gains of considering the information on the within-unit correlation to bound the between-unit correlation. Of course, if we have information on all relevant distance metrics in the cross section, then it is possible to correct for spatial correlation with a more powerful test, as the correctly specified parametric bootstrap shows. Still, these simulations illustrate that it is possible to construct valid tests even when such information is unavailable or the researcher is not willing to impose a structure on the spatial correlation.

\section{Empirical Illustration} \label{illustration}

We also illustrate our findings analyzing the effects of the Massachusetts 2006 health care reform. This reform was analyzed by \cite{Sommers} using a DID design comparing 14 Massachusetts counties with  513 control counties from 45 different states that were selected based on a propensity score to be more similar with the treated counties (we find similar results if we consider a DID regression using all counties, so that there is no pre-selection of control counties). \cite{Sommers} find a reduction of 2.9\%-4.2\% in mortality in Massachusetts relative to the controls after the reform (depending on whether covariates are included). 

As shown in Figure 1 from \cite{Sommers}, the outcome variables in the pre-treatment periods followed parallel trajectories when we compare Massachusetts and the control groups. This provides some evidence in favor of Assumption \ref{As_id}, although it does not guarantee that. \cite{Sommers} relied on standard errors clustered at the state level, which does not work well in this setting with a single treated state.  Their inference procedures were then re-analyzed by  \cite{Kaestner}, who considered permutation tests at the county level. While permutation tests are usually considered in a design-based framework, if we consider this approach in our framework, it is  asymptotically equivalent to CT at the county level {(see \cite{alvarez2025inferencetreatedunits})}. Therefore, this would also be problematic if counties within the same state are spatially correlated, as we show in Section \ref{Problems}.\footnote{
To provide evidence on the presence of spatial correlation in time-varying unobservables across counties within the same sate, we consider a ``design-based'' placebo exercise \citep{ferman2026usedesignbasedsimulations}. More specifically,  we exclude Massachusetts from our sample and consider an exercise where we randomly generate placebo treatment assignments in 2007 for half of the remaining states in the sample. For each one of these placebo assignments, we calculate the DID estimator and test the null of no effect using a $t$-statistic using clustered standard errors at the county level at 5\% significance level. Following \cite{ferman2026usedesignbasedsimulations}, we should expect rejection rates around 5\% if there is no spatial correlation across counties, but larger rejection rates if errors are spatially correlated. We find a rejection rate of 62\% for all-cause mortality, and of 54\% health care-amenable shocks, thus evidencing the presence of state-level spatial correlation.

}

We note that this is a setting in which (i) we have only a single treated state, (ii) errors are likely correlated across counties within the same state, and (iii) there is variation in population sizes across counties/states that may lead to heteroskedasticity. If we assume that errors are independent across states, and that heteroskedasticity depends only on state sizes, 
then FP at the state level would be an appropriate inference method in this setting. Moreover, given Corollary \ref{Cor_N1}, this inference method remains valid even if we assume that state-level errors are weakly dependent, since in this case we have only a single treated state. We consider, therefore, p-values and CIs from FP at the state-level as a benchmark. As presented in Table \ref{Table_Empirical}, in this case we fail to reject the null of no effect, for the two outcomes we considered.

While in this application we have a well-defined distance measure for the spatial correlation (which, in this case, is the information on the states),  we consider a ``thought experiment'' in which the applied researcher does not have such information, in order to  illustrate our main results. The main advantage is that, in this case, we can contrast our findings with the conclusions based on FP at the state level, which we use as a benchmark. Table \ref{Table_Empirical} thus reports the p-values, as well as the length of the resulting 95\% confidence interval obtained through test inversion, of alternative inference methods at the county level.

First, note that we would reject the null at the 5\% level if we relied on most of the inference methods that ignore spatial correlation (FP, CRVE and WCB at county level), contrasting with our benchmark results using FP at the state level. This happens because, as discussed in Section \ref{Problems}, spatial correlation among the treated units leads to over-rejection. The only exception is CT at the county level. This happens because, while spatial correlation induces over-rejection in this case, the fact that treated counties are relatively larger induces under-rejections. 

We also consider how our conservative tests would perform in this case (again, considering this ``thought experiment'' in which information on states is unavailable).   The conservative tests given by Methods 3 (Section \ref{Sec_cons1}) and  4 (Section \ref{Sec_cons2}), present p-values similar to the ones from FP at the state level (our benchmark). When we consider Method 4, the  length of the 95\% confidence interval is 48\% larger for all cause mortality, and 28\% larger for healthcare amenable mortality. Therefore, there is some loss in terms of power, but in settings in which a distance metric is not available, this may be a cost we would have to pay to provide a test that controls for size in the presence of spatial correlation. Moreover, Methods 3 and 4 offer clear improvements over Method 1 (Section \ref{bounds_arbitrary}). Indeed, this procedure leads to confidence intervals with around double the length than than FP at the state level. We also note that, in this particular application, Method 2 (Section \ref{Bounds_positive}) yields identical conclusions to Method 1, as both inference procedures collapse to the same test function given the larger length of the confidence intervals obtained from inversion of the \cite{Benjamini1995} procedure.

Overall, our results indicate that Methods 3 and 4 would perform well even if we DID not have information on a distance metric, and despite the fact that we have spatial correlation across counties.

\begin{table}[h]

\caption{{\bf Alternative inference methods for \cite{Sommers}}} \label{Table_Empirical}
\resizebox{\textwidth}{!}{
\begin{tabular}{lccccc}
\hline
\hline

& \multicolumn{2}{c}{All cause mortality} & &  \multicolumn{2}{c}{Health care-amenable mortality} \\
& \multicolumn{2}{c}{(deaths per 100,000 adults)} & &  \multicolumn{2}{c}{(deaths per 100,000 adults)} \\ \cline{2-3} \cline{5-6}

& p-value & length of 95\% CI && p-value & length of 95\% CI \\

& (1) & (2) & & (3) & (4) \\

\hline

CT and FP at state level \\

~~~CT & 0.265 & 123.66 &  & 0.221 & 72.14 \\
 
~~~ FP (benchmark) & 0.202 & 52.23 &  & 0.133 & 28.54 \\
 
CRVE and WCB at county level \\
~~~ CRVE & 0.002 & 16.08 &  & 0.001 & 11.36 \\
 
~~~ WCB & 0.046 & 20.53 &  & 0.028 & 14.36 \\
 
CT and FP at county level \\

~~~ CT & 0.087 & 36.09 &  & 0.073 & 26.24 \\
 
~~~ FP & 0.010 & 27.10 &  & 0.001 & 14.41 \\
 
Conservative tests \\

~~~ Method 1 & 0.736 & 106.70 &  & 0.554 & 56.92 \\

~~~ Method 2 & 0.736 & 106.70 &  & 0.554 & 56.92 \\

~~~ Method 3 & 0.247 & 83.08 &  & 0.187 & 43.31 \\
 
~~~ Method 4 & 0.239 & 77.31 &  & 0.126 & 36.54 \\

\hline

\end{tabular}
}
     
\singlespacing\footnotesize
\noindent \textit{Notes:} This table presents p-values and lengths of confidence intervals for a series of alternative inference methods for the application from \cite{Sommers}.  We consider a DID estimator based on a TWFE regression with no covariates. Point estimates are slightly different than reported in the original paper because we use the publicly available data set (so we do not have death counts for cells with fewer than 10 deaths), and because we weight observations by population mean across years. We also restrict to counties with non-missing information for all years. We end up with 485 control counties, as compared to 513 from the original study. Point estimates are $-12.77$ for all cause mortality and $-9.94$ for health care-amenable mortality.       
\end{table}

\section{Conclusion}

We consider the problem of inference in DID  when there are few treated units and errors are spatially correlated. We first show that, when there is a single  treated unit, the main inference methods proposed by CT and FP, which were  designed for settings with few treated and many control units, remain asymptotically valid when errors are weakly dependent {(even when the relevant distance metric is unspecified)}. This extends the set of possible applications in which the tests proposed by CT and FP can be reliably used when there is only a single treated unit. However, these methods can lead to over-rejection with more than one treated unit. We propose a series of alternative inference methods that are asymptotically valid, though generally conservative, in the presence of spatial correlation. These tests provide  interesting alternatives when spatial correlation is likely relevant, but the researcher does not have information on a distance metric or is not willing to impose a structure on the spatial correlation. {By providing a menu of alternative inference methods, applied researchers can evaluate in their empirical application the trade-offs around the assumptions they are willing to make about the spatial correlation and the length of the confidence intervals.  }

\section*{Acknowledgments}

 We would like to thank Jon Roth and Chris Taber for comments and suggestions.  Lucas Barros provided exceptional  research assistance. We also thank Benjamin Sommers for useful discussions and for providing the county FIPS codes for the counties used in \cite{Sommers}. Bruno Ferman gratefully acknowledges financial support from FAPESP and CNPq.


\singlespacing

\bibliographystyle{apalike}
\bibliography{bib.bib}

\end{document}


\pagestyle{plain}
 \setcounter{table}{0}
\renewcommand\thetable{\thesection.\arabic{table}}

\setcounter{figure}{0}
\renewcommand\thefigure{\thesection.\arabic{figure}}
\maketitle

\tableofcontents

\appendix

\onehalfspacing

\section{Proof of results in the main text} 
This Appendix presents the proofs of the main results of the paper.

\subsection{Proof of Proposition \ref{Prop_N1_FP}}
\label{Proof_N1}
\begin{proof}
We begin by showing that $\hat{F}_\xi(c) \overset{p}{\to} F_\xi(c)$ for every $c \in \mathbb{R}$. First, note that, by the properties of the TWFE estimator, $\hat{W}_i = W_i - \frac{1}{N_0}\sum_{s \in \mathcal{I}_0} W_s$ for $i \in \mathcal{I}_0$. Next, define the random variable $i^* \sim \operatorname{Uniform}(\mathcal{I}_0)$, independently from the data. Observe that:

\begin{equation}
\label{eq_up_bound_star}
|\hat{\xi}_{i^* }- \xi_{i^*}| \leq \max_{j \in \mathcal{I}_0}\left|\frac{1}{h(Z_j;\hat \delta)}\right|\left|\frac{1}{N_0} \sum_{s \in \mathcal{I}_0} W_s\right| + \max_{j \in \mathcal{I}_0}\left|\frac{h(Z_j;\delta)}{h(Z_j;\hat \delta)} - 1\right||\xi_{i^*}|\, .
\end{equation}

Now we fix $\epsilon \in (0, \underline{h})$ and notice that, on the event that $\max_{s\in \mathcal
{I}_0}| h(Z_s;\widehat \delta) - h(Z_s; \delta)|\leq \epsilon$, Assumption \ref{As_W}.(ii) implies that:

$$\max_{j \in \mathcal{I}_0}\left|\frac{h(Z_j;\delta)}{h(Z_j;\hat \delta)} - 1\right| \leq \frac{\epsilon}{\underline{h}-\epsilon}\, ,$$
Consequently, since, by Assumption \ref{As_W}.(iv), $\mathbb{P}[\max_{s\in \mathcal
{I}_0}| h(Z_s;\widehat \delta) - h(Z_s; \delta)|> \epsilon]\to 0$ as $N_0 \to \infty$, one concludes that, as $N_0 \to \infty$:

$$\mathbb{P}\left[\max_{j \in \mathcal{I}_0}\left|\frac{h(Z_j;\delta)}{h(Z_j;\hat \delta)} - 1\right| > \frac{\epsilon}{\underline{h}-\epsilon}\right] \to 0\, ,$$
and given that $\epsilon$ can be chosen arbitrarily small, we conclude that

$$\max_{j \in \mathcal{I}_0}\left|\frac{h(Z_j;\delta)}{h(Z_j;\hat \delta)} - 1\right|  = o_{\mathbb{P}}(1)\, .$$

Next, notice that, under Assumption \ref{As_W}.(i), $\mathbb{E}[|\xi_{i^*}|]=\mathbb{E}[|\xi|]<\infty$, from which we infer that $|\xi_{i^*}|=o_{\mathbb{P}}(1)$. Combining these facts back on  \eqref{eq_up_bound_star}, we conclude that: 

$$|\hat{\xi}_{i^* }- \xi_{i^*}| = o_{\mathbb{P}}(1) \, .$$

Next, define $\mathcal{A} = \sigma(\{Y_{s1},...,Y_{sT},Z_s\}_{ \mathcal{I}_1 \cup \mathcal{I}_0})$, and note that, by Markov inequality, for any $\epsilon,\delta>0$:

$$\mathbb{P}[\mathbb{P}[|\hat{\xi}_{i^* }- \xi_{i^*}|>\epsilon|\mathcal{A}]>\delta] \leq \frac{\mathbb{P}[|\hat{\xi}_{i^* }- \xi_{i^*}|>\epsilon]}{\delta}\to 0 \, ,$$
from which we conclude that $|\hat{\xi}_{i^* }- \xi_{i^*}|=o_{\mathbb{P}^*}(1)$ in probability, where $\mathbb{P}^*$ denotes the conditional law $\mathbb{P}[\cdot|\mathcal{A}]$. Finally, we note that, for any $l \in \mathbb{R}$, Assumption \ref{As_W}.(ii) implies that:

$$\mathbb{P}[\xi_{i^*}\leq l|\mathcal{A}] = \frac{1}{N_0} \sum_{s \in \mathcal{I}_0}  \mathbbm{1}\{ \xi_s \leq l \}  \buildrel p \over \rightarrow  F_\xi(l)\, ,$$
and the desired conclusion then follows from the observation that, for a given $c \in \mathbb{R}$ and every $\nu > 0$:

\begin{equation}
    \begin{aligned}
|\hat{F}_\xi(c) - F_\xi(c)| = |\mathbb{P}[\hat \xi_{i^*}\leq c|\mathcal{A}]   - F_\xi(c)|\leq  \\|\mathbb{P}[\hat \xi_{i^*}\leq c|\mathcal{A}] - \mathbb{P}[ \xi_{i^*}\leq c|\mathcal{A}]| + |\mathbb{P}[\xi_{i^*}\leq c|\mathcal{A}] - F_\xi(c)| \leq \\
        \mathbb{P}[|\hat{\xi}_{i^*}-\xi_{i^*}|>\nu|\mathcal{A}] + \mathbb{P}[c-\nu \leq \xi_{i^*}\leq c+\nu] + |\mathbb{P}[\xi_{i^*}\leq c|\mathcal{A}] - F_\xi(c)| \overset{p}{\to} F_{\xi}(c+\nu)-F_\xi(c-\nu)
    \end{aligned}
\end{equation}
from which it follows that, for any $\epsilon,\nu>0$, $\mathbb{P}[|\hat{F}_\xi(c) - F_\xi(c)|>F_{\xi}(c+\nu)-F_\xi(c-\nu) + \epsilon ]\to 0$. Since $F_\xi$ is continuous on $c$, we conclude that $|\hat{F}_\xi(c) - F_\xi(c)| =o_{\mathbb{P}}(1)$, thus establishing pointwise convergence of $\hat{F}_\xi$.
 
 Finally, since the choice of $c$ was arbitrary and $F_\xi$ is a continuous distribution function, it follows that the convergence is uniform, i.e. $\sup_{c \in \mathbb{R}}|\hat{F}_\xi(c) - F_\xi(c)| = o_{\mathbb{P}}(1)$ \citep[page 339]{Vaart1998}.
\end{proof}

\subsection{Proof of Corollary \ref{Cor_N1}}
\label{proof_Cor}
\begin{proof}
Observe that, under the null $\frac{\hat \alpha - \alpha_0}{h(Z_1;\hat \delta)}\overset{p}{\to}, \xi_1$. Consequently, Proposition \ref{Proof_N1} entails that: 

$$\hat{F}_\xi\left(\frac{\hat{\alpha}-\alpha_0}{h(Z_1;\hat{\delta
})}\right) \overset{p}{\to}F_\xi(\xi_1)\,,$$
where we have that $F_\xi(\xi_1) \sim \operatorname{Uniform}[0,1]$ due to continuity of $F_\xi$. The conclusion then follows by noticing, first, that by the continuous mapping theorem:

\begin{equation}
    \begin{aligned}
    \phi_{\text{FP}} = \mathbbm{1}\left\{\hat{F}_\xi\left(\frac{\hat{\alpha}-\alpha_0}{h(Z_1;\hat{\delta
})}\right) <\tau/2\right\} +   \mathbbm{1}\left\{\hat{F}_\xi\left(\frac{\hat{\alpha}-\alpha_0}{h(Z_1;\hat{\delta
)}}\right) > 1- \tau/2\right\} \overset{p}{\to} \\ \mathbbm{1}\left\{{F}_\xi\left(\frac{\hat{\alpha}-\alpha_0}{h(Z_1;\hat{\delta
})}\right) <\tau/2\right\} +   \mathbbm{1}\left\{{F}_\xi\left(\frac{\hat{\alpha}-\alpha_0}{h(Z_1;\hat{\delta
)}}\right) > 1- \tau/2\right\} \, ,
    \end{aligned} 
\end{equation}
and, consequently, by the bounded convergence theorem:

\begin{equation}
    \begin{aligned}
    \mathbb{E}[\phi_{\text{FP}}] \to \mathbb{P}\left[{F}_\xi\left(\frac{\hat{\alpha}-\alpha_0}{h(Z_1;\hat{\delta
})}\right) <\tau/2\right] +   \mathbb{P}\left[{F}_\xi\left(\frac{\hat{\alpha}-\alpha_0}{h(Z_1;\hat{\delta
)}}\right) > 1- \tau/2\right] = \tau
    \end{aligned} \, .
\end{equation}
\end{proof}
\subsection{Proof of  Proposition \ref{prop_es}}
\label{proof_prop_es}
\begin{proof}
  In this case, it suffices to show that, as $N_0\to \infty$:
  
  $$\hat U \overset{p}{\to}  \sum_{i=1}^{N_1} \frac{1}{N_1}h(Z_i; \delta) \times \min\left\{ {Q}_\xi\left(1-\frac{\tau}{2N_1}\right), \frac{1}{\tau/2}\int_{ Q_\xi\left(1-\frac{\tau}{2}\right)}^\infty s  F_\xi(ds) \right\}\, ,$$
  and
    $$ \hat L \overset{p}{\to} \sum_{i=1}^{N_1} \frac{1}{N_1}h(Z_i; \delta) \times \max\left\{ {Q}_\xi\left(\frac{\tau}{2N_1}\right), \frac{1}{\tau/2}\int_{-\infty}^{ Q_\xi\left(\frac{\tau}{2}\right)}s  F_\xi(ds) \right\}\, ,$$
 as the conclusion then follows by a similar argument to the proof of Corollary \ref{Cor_N1}. 
    
    We analyze consistency of $\hat U$, as the case of $\hat L$ is analogous and thus omitted. We observe that, as a consequence of Proposition \ref{Prop_N1_FP} and Lemma \ref{lemma_uniform_qt}, as $N_0 \to \infty$, $\hat Q_\xi(u) \overset{p}{\to} Q(u)$ for any $u \in (0,1)$. Consequently:
  $$\hat{Q}_\xi\left(1-\frac{\tau}{2N_1} \right) \overset{p}{\to} {Q}_\xi\left(1-\frac{\tau}{2N_1}\right)\, .$$

  Next, let $\tilde{F}_\xi$ denote the (unfeasible) empirical distribution of the $\xi_i$, $i \in \mathcal{I}_0$, and $\tilde{Q}_\xi$ the associated quantile function. We note that:

  $$\left|\frac{1}{\tau/2}\int_{\hat Q_\xi\left(1-\frac{\tau}{2}\right)}^\infty s \hat F_\xi(ds) -\frac{1}{\tau/2}\int_{\tilde Q_\xi\left(1-\frac{\tau}{2}\right)}^\infty s \tilde F_\xi(ds)\right|\leq \left| \max_{i \in \mathcal{I}_0}\frac{h(Z_i; \hat \delta) - h(Z_i; \delta)}{h(Z_i; \hat \delta) } \right| \frac{1}{N_0}\sum_{i \in \mathcal{I}_0} |\xi_i|\, .$$

  Since, $\left| \max_{i \in \mathcal{I}_0}\frac{h(Z_i; \hat \delta) - h(Z_i; \delta)}{h(Z_i; \hat \delta) } \right|  = o_p(1)$ and $ \frac{1}{N_0}\sum_{i \in \mathcal{I}_0} |\xi_i| = O_P(1)$ by Assumption \ref{As_W}, we conclude that:

  $$\left|\frac{1}{\tau/2}\int_{\hat Q_\xi\left(1-\frac{\tau}{2}\right)}^\infty s \hat F_\xi(ds) -\frac{1}{\tau/2}\int_{\tilde Q_\xi\left(1-\frac{\tau}{2}\right)}^\infty s \tilde F_\xi(ds)\right| = o_p(1)\, .$$

  Finally, we note that, by Lemma \ref{lemma_ES_CONVERGENCE}:

  $$\frac{1}{\tau/2}\int_{\tilde Q_\xi\left(1-\frac{\tau}{2}\right)}^\infty s \tilde F_\xi(ds) \overset{p}{\to} \frac{1}{\tau/2}\int_{ Q_\xi\left(1-\frac{\tau}{2}\right)}^\infty s  F_\xi(ds) \, .$$
  and that $\hat{U} \overset{p}{\to} U$ then follows by the continuous mapping theorem.
  \end{proof}

  \subsection{Proof of Proposition \ref{prop_mht}}
\label{proof_mht}

\begin{proof}
We begin by showing that, as $N_0 \to \infty$, $c^* \overset{p}{\to} \sum_{i=1}^{N_1}\frac{1}{N_1}h(Z_{(N_1+1-i)};{\delta}) {Q}_{|\xi|}\left(1-\frac{\tau}{N_1}\right)$. Observe that, for each $x \in \mathbb{R}$:

$$\left|\hat{F}_{|\xi|}(x) - F_{|\xi|}(x)\right|\leq |\hat{F}_{\xi}(x) -{F}_{\xi}(x) | +|\hat{F}_{\xi}(-x) -{F}_{\xi}(-x) | + \frac{1}{N_0}\sum_{i \in \mathcal{I}_0} \mathbbm{1}\{\hat \xi_i = -x \}$$

By Propositon \ref{Prop_N1_FP}, $\sup_{x \in \mathbb{R}}|\hat{F}_{\xi}(x) -{F}_{\xi}(x) | = o_{p}(1)$ and $\sup_{x \in \mathbb{R}}|\hat{F}_{\xi}(-x) -{F}_{\xi}(-x) | = o_{p}(1)$. Moreover, we notice that, by the uniform convergence of Proposition \ref{Prop_N1_FP} and continuity of $F_\xi$, we have that:
\begin{equation*}
    \begin{aligned}
        \sup_{x \in \mathbb{R}} \frac{1}{N_0}\sum_{i \in \mathcal{I}_0} \mathbbm{1}\{\hat \xi_i = -x \}  = \sup_{x \in \mathbb{R}} \left(\frac{1}{N_0}\sum_{i \in \mathcal{I}_0} \mathbbm{1}\{\hat \xi_i \leq x \} - \lim_{k \to \infty}   \frac{1}{N_0}\sum_{i \in \mathcal{I}_0} \mathbbm{1}\{\hat \xi_i \leq x - k^{-1} \}\right) = \\
        \sup_{x \in \mathbb{R}} \left[\left(\frac{1}{N_0}\sum_{i \in \mathcal{I}_0} \mathbbm{1}\{\hat \xi_i \leq x \} - F_\xi(x) \right) - \lim_{k \to \infty} \left(   \frac{1}{N_0}\sum_{i \in \mathcal{I}_0} \mathbbm{1}\{\hat \xi_i \leq x - k^{-1} \} - F_\xi(x - k^{-1})\right)\right] \leq \\ 2 \sup_{x \in \mathbb{R}}|\hat{F}_{\xi}(x) -{F}_{\xi}(x) | = o_{p}(1)
    \end{aligned}\, ,
\end{equation*}
thus implying that $\sup_{x \in \mathbb{R}} \frac{1}{N_0}\sum_{i \in \mathcal{I}_0} \mathbbm{1}\{\hat \xi_i = -x \} =o_p(1)$. Consequently, $\sup_{x \in \mathbb{R}} \left|\hat{F}_{|\xi|}(x) - F_{|\xi|}(x)\right|=o_p(1)$, and Lemma \ref{lemma_uniform_qt} yields that $\hat{Q}_{|\xi|}(u) \overset{p}{\to} {Q}_{|\xi|}(u) $ for any $u \in (0,1)$. We thus obtain that  $c^* \overset{p}{\to} \sum_{i=1}^{N_1}\frac{1}{N_1}h(Z_{(N_1+1-i)};{\delta}) {Q}_{|\xi|}\left(1-\frac{\tau}{N_1}\right)$, as desired. Next, let $\boldsymbol{c}^*$ denote the probability limit of $c^*$, and $\boldsymbol{U}$ and $\boldsymbol{L}$ denote respectively the probability limits of $\hat{U}$ and $\hat{L}$ -- recall the latter exist by the proofs of Propositions \ref{prop_makarov} (for the Makarov bound) and \ref{prop_es} (in the case of the union \textit{cum} expected shortfall bound).  We then have that, by consistency:

$$\lim_{N_0 \to \infty}\mathbb{P}[\{2c^*> \hat{U} - \hat{L}\}] = \begin{cases}
    1 & \text{if } 2\boldsymbol{c}^* > \boldsymbol{U}-\boldsymbol{L}\\
        0 & \text{if } 2\boldsymbol{c}^* \leq \boldsymbol{U}-\boldsymbol{L}
\end{cases}\, .$$

Consequently, we have that:

$$\limsup_{N_0 \to \infty}\mathbb{P}[\alpha \notin \mathcal{I}_\text{refined}] = \begin{cases}
    \limsup_{N_0 \to \infty}\mathbb{P}[\alpha \notin [\hat{\alpha} - \boldsymbol{U}, \hat{\alpha} - \boldsymbol{L}] ] & \text{if } 2\boldsymbol{c}^* > \boldsymbol{U}-\boldsymbol{L}\\
         \limsup_{N_0 \to \infty}\mathbb{P}[\alpha \notin \mathcal{I}_{BH}]  & \text{if } 2\boldsymbol{c}^* \leq \boldsymbol{U}-\boldsymbol{L}
\end{cases}$$

That $ \limsup_{N_0 \to \infty}\mathbb{P}[\alpha \notin [\hat{\alpha} - \boldsymbol{U}, \hat{\alpha} - \boldsymbol{L}] ] \leq \tau$ follows from  Proposition \ref{prop_makarov} (for the Makarov bound) and  \ref{prop_es} (jn the case of the union \textit{cum} expected shortfall bound). That $  \limsup_{N_0 \to \infty}\mathbb{P}[\alpha \notin \mathcal{I}_{BH}]\leq$ follows from observing that, by the continuous mapping theorem, $\mathbbm{1}\{ \alpha \notin  \mathcal{I}_{BH}\} \overset{p}{\to} \mathbbm{1}\{\alpha \notin [\alpha + \frac{1}{N_1}\sum_{i \in \mathcal{I}_1} W_i - \boldsymbol{c}^*, \alpha + \frac{1}{N_1}\sum_{i \in \mathcal{I}_1} W_i + \boldsymbol{c}^* \tilde]\}$, and, therefore, by the continuous mapping theorem

$$\mathbb{P}[\alpha \notin \mathcal{I}_{BH}] \to \mathbb{P}\left[\left\{\alpha \notin \left[\alpha + \frac{1}{N_1}\sum_{i \in \mathcal{I}_1} W_i - \boldsymbol{c}^*, \alpha + \frac{1}{N_1}\sum_{i \in \mathcal{I}_1} W_i + \boldsymbol{c}^* \right]\right\}\right]\leq \tau\, ,$$
where the last inequality follows from Theorem 1.2 of \cite{Benjamini2001}.
\end{proof}

\section{Sharper bounds for Method 1 when $N_1=2$}
\label{sec_makarov}

When $N_1=2$, there exists an upper bound to $Q_{\widetilde{W}}$ that is agnostic about the dependence structure, easy to compute, and generally sharper than those considered by the version of Method 1 discussed in the main text (Algorithm \ref{alg_es}). Specifically, for a given $u \in (0,1)$, \cite{Makarov1982} yields the following bound:

$$Q_{\widetilde{W}}(u) \leq \inf_{p\in[u,1]} \left\{ \frac{1}{2}h(Z_1;\delta) Q_{\xi}(p) +  \frac{1}{2}h(Z_2;\delta) Q_{\xi}(1-p+u) \right\} =: \bar{Q}_{\widetilde{W}}(u) \, . $$

This bound is known to be pointwise-optimal, in the sense that, for each $\epsilon > 0$, there is a joint law $\mathbb{G}$ of $(\xi_1,\xi_2)$ whose marginals coincide with $F$ such that $1-u \geq \mathbb{G}[\widetilde{W}> \bar{Q}_{\widetilde{W}}(u)]\geq 1-u - \epsilon $ \citep[Theorem 13]{zhang2025boundsdistributionsumrandom}.

To construct a test of the null $H_0: \alpha = \alpha_0$ against the two-sided alternative $H_1: \alpha \neq \alpha_0$ at the $\tau$-significance level, one must also find a lower bound to $Q_{\widetilde{W}}(\tau/2)$. This can be done by applying the \cite{Makarov1982} upper bound to $Q_{-\widetilde{W}}(1-\tau/2)$. Since computation of the program defining the Makarov bounds requires knowledge of the quantile function $Q_\xi$, we propose that computation be performed by replacing these with the empirical quantiles $\hat{Q}_\xi$ obtained from $\hat{F}_\xi$, which produces the following algorithm for testing the null:

\renewcommand{\thealgocf}{M1'}
\begin{algorithm} [h]
Run the DID estimator, and store $\widehat \alpha$ and the residuals  $\widehat W_s = \frac{1}{T-t^\ast} \sum_{t \in \mathcal{T}_1}\hat \eta_{st} -  \frac{1}{t^\ast} \sum_{t \in \mathcal{T}_0} \hat \eta_{st}$ for all $s \in \mathcal{I}_0$ \;
 Estimate $\delta$ using the residuals from the controls, and compute the normalized residuals $\widehat \xi_s = \widehat W_s / h(Z_s,\widehat \delta)$, $s \in \mathcal{I}_0$ \;
 Compute the empirical distribution of the normalized residuals $\hat{F}_\xi$ and the associated empirical quantile functions $\hat{Q}_\xi$ \;
Estimate the bounds:
$$ \hat U = \inf_{p\in [1-\tau/2,1]} \left\{ \frac{1}{2}h(Z_1;\hat \delta) \hat Q_{\xi}(p)+ \frac{1}{2}h(Z_2;\hat \delta) \hat Q_{\xi}(1 -p + (1-\tau/2)) \right\}$$
$$ \hat L = \sup_{p\in [0,\tau/2]}  \left\{\frac{1}{2}h(Z_1;\hat \delta) \hat Q_{\xi}(p)+ \frac{1}{2}h(Z_2;\hat \delta) \hat Q_{\xi}(\tau/2 - p)) \right\}$$

 Reject the null if $\hat \alpha - \alpha_0> \hat U$ or $\hat \alpha - \alpha_0< \hat L$.
	\caption{Test based on \cite{Makarov1982} bounds}
     \label{alg_makarov}
\end{algorithm}

The following proposition establishes asymptotic validity of the above approach.

\begin{proposition}
\label{prop_makarov}
    Suppose that $N_1=2$, and that Assumptions \ref{As_id} and \ref{As_W} hold. Let ${\phi}_{1'}$ denote the decision rule from Algorithm \ref{alg_makarov}. Then, under the null, as $N_0 \to \infty$, $\limsup_{N_0\to \infty}\mathbb{E}[\phi_{1'}]\leq \tau $. 
    \begin{proof}
        See Appendix \ref{proof_makarov}.
    \end{proof}
\end{proposition}

\paragraph{Connection to the union bound} Notice that the upper bound $\bar{Q}(u)$ is always no greater than an upper bound given by the union bound, i.e.\footnote{The bound follows from the observation that, for random variables $X$ and $Y$, and any $u \in (0,1)$: $$\mathbb{P}\left[X+Y> Q_{X}\left(\frac{1+u}{2}\right) + Q_{Y}\left(\frac{1+u}{2}\right)\right] \leq \mathbb{P}\left[X> Q_{X}\left(\frac{1+u}{2}\right) \right]+ \mathbb{P}\left[Y>Q_{Y}\left(\frac{1+u}{2}\right)\right]  \leq 1-u \, .$$}

$$\bar{Q}_{\widetilde{W}}(u) \leq  \frac{1}{2}h(Z_1;\delta) Q_{\xi}\left(\frac{1+u}{2}\right) +  \frac{1}{2}h(Z_2;\delta) Q_{\xi}\left(\frac{1+u}{2}\right)\, .$$

The following lemma provides a set of sufficient conditions for both bounds to coincide, i.e. for the union bound correction to yield the sharpest upper bound to the $u$-quantile of $\widetilde{W}$ under arbitrary dependence.  We observe that a similar result appears as Proposition 2 of \cite{Embrechets2013}.

\begin{lemma}
\label{lemma_union bound}
    Consider the case $N_1=2$. In the homoskedastic setting, i.e. $\sigma = h_1(Z_1;\delta)=h_2(Z_2;\delta)$, if the following conditions hold:
    \begin{enumerate}
    \item[(i)] The support of $F$ contains the interval $(Q_F(u),\infty)$.
    \item[(ii)] $F$ is differentiable on $(Q_F(u),\infty)$, and the the corresponding density $f$ is strictly monotone on $(Q_F(u),\infty)$.
\end{enumerate}
then $\bar{Q}_{\widetilde{W}}(u) = \sigma Q_{\xi}\left(\frac{1+u}{2}\right)$.
\begin{proof}
    See Appendix \ref{proof_union_sharp}.
\end{proof}
\end{lemma}

\section{Construction of confidence intervals based on  the \cite{Benjamini1995} correction}
\label{Appendix_MHT}
The procedure proposed by \cite{Benjamini1995} for controlling the FDR is as follows. For testing the nulls $\alpha_s = a_s$, $s \in \mathcal{I}_1$, with control of the FDR at level $\tau$, we first compute $N_1$ p-values for the tests, which in our setting are given by:

$$\hat{p}_s = 1 - \hat{F}_{|\hat{\xi}|}\left(\left|\frac{\hat{\alpha}_s - a_s}{h(Z_s;\hat \delta)}\right|\right) \, ,$$
where $\hat{\alpha}_s$ is the DID estimator that only uses treated unit $s$ and the controls. We then order the nulls and corresponding p-values from smallest to largest, $\hat p_{(1)} \leq \hat p_{(2)} \leq ... \hat p_{(N_1)}$, and compute: 

$$k = \max \left\{ j \leq N_1:  \hat{p}_{(j)} < \frac{j}{N_1} \tau \right\} \,.$$

We then reject the first $k$ nulls. If no $k$ satisfies the above requirement, we do not reject \emph{any} null.

From the above description, it is clear that the set of null hypotheses for which no elements are rejected must satisfy $\hat{p}_{(j)} \geq \frac{j}{N_1}\tau$ for all $j \leq N_1$. Let $\mathcal{P}_{N_1}$ denote the set of permutations on $\{1,\ldots, N_1\}$. It follows that the confidence set for heterogeneous effects is given by:

\begin{equation*}
	\begin{aligned}
		I_{BH}((\alpha_s)_{s \in \mathcal{I}_1}; 1-\tau) = {\bigcup}_{\rho \in \mathcal{P}_{N_1}} \prod_{j=1}^{N_1} [\hat{\alpha}_{\rho(j)} - \hat{q}_{|\xi|}(1-\rho(j)\tau/N_1), \hat{\alpha}_{\rho(j)} + \hat{q}_{|\xi|}(1-\rho(j)\tau/N_1)]  \, .
	\end{aligned}
\end{equation*}

Since, for each $\rho \in \mathcal{P}_{N_1}$,  the set inside the outermost union is connected, it follows that its projection is an interval. A simple argument reveals that the union of these intervals as $\rho$ ranges $\mathcal{P}_{N_1}$ is also an interval, which establishes that the confidence set for a weighted average of effects $\alpha_w = \sum_{s \in \mathcal{I}_1} \omega_s \alpha_s$ is given by:

$$I_{BH}(\alpha_w; 1-\alpha) =  \left[\hat \alpha_{\omega} - c^*,  \hat \alpha_{\omega} + c^*\right] \, ,$$
where $c^* = \max_{\rho \in \mathcal{P}_{N_1}} \sum_{j=1}^{N_1} \omega_{\rho(j)} h(Z_{\rho(j)}; \hat \delta) \hat q_{|\xi|}(1- \frac{j}{N_1}\tau)$. The constant $c^*$ can be quickly computed by ordering $\hat m_s := \omega_{s}  h(Z_{s}; \hat \delta)$ from \textbf{largest to smallest}, $\hat m_{(1)} \geq \hat m_{m(2)} \geq \ldots \geq \hat m_{(N_1)}$ and noting that $c^* = \sum_{j=1}^{N_1} \hat m_{(j)} \hat q_{|\xi|}(1- \frac{j}{N_1}\tau)$.

\section{Ranking confidence intervals obtained from inversion of conservative Methods 1, 2, 3 and 4}
\label{app_rank}
In the main text, we have discussed four methods for inference on average treatment effects that, upon test inversion, may be used to construct confidence intervals\footnote{Or prediction intervals, if treatment effects are stochastic, as discussed in Remark \ref{Remark_AF} in the main text.} whose validity rests on different set of assumptions. At one extreme, we can construct confidence intervals based on bounds on the quantile function of sums of random variables without imposing any assumptions on spatial dependence (Section \ref{bounds_arbitrary}). These methods remain valid irrespective of the true dependence structure of the errors. At the other extreme, the methods given \ref{Sec_cons1} and \ref{Sec_cons2} rely on the assumption that the applied researcher can point to a particular dependence structure that would constitute the worst-case scenario for serial dependence (in the first case, the comonotone copula that captures perfect positive dependence between units; in the second case, a hypothetical scenario where all units lie within the same cluster). As an intermediate approach, the method in Section \ref{Bounds_positive} does not specify a worst-case scenario for spatial correlation, but restricts the dependence structure of the errors to be positive, in the sense of Assumption \ref{ass_psrd}. 

Even though the assumptions underlying Methods 2 to 4 are generally not nested, the proposition below shows that, at least in the case where the empirical distribution of the residuals $\hat \xi_s$ is symmetric, confidence intervals obtained by inversion of the corresponding test procedures can be ranked with respect to the inclusion order. For example, one can show that, in the weighted DID case, the confidence sets of Method 1 include those of Method 2, those of Method 2 include Method 3, and Method 3 includes Method 4. This reveals that being less specific about the worst-case scenario for spatial correlation, as Methods 1 and 2 are, naturally translates into larger confidence sets than those implied when one is willing to point to a particular worst-case scenario -- as Methods 3 and 4 do. This pattern generates a natural trade-off between the plausibility of the assumptions required by each method and the resulting power of the inference method.

In what follows, let $\mathcal{I}_j$ denote the resulting confidence interval for methods $j=1,\ldots, 4$.
\begin{proposition}
\label{prop_inclusion}
    Consider the unweighted DID estimator, inference based on a common parametric model for heteroskedasticity ($\mathbb{V}[Z_s] = h(Z_s;\delta)$)  and the resulting confidence interval obtained from inverting test procedures 1 to 3.  We then have that $\mathcal{I}_1 \supseteq \mathcal{I}_3$ and $\operatorname{length}(\mathcal{I}_1)\geq  \max\{\operatorname{length}(\mathcal{I}_2),   \operatorname{length}(\mathcal{I}_3)\}$. If the empirical distribution of the error term is symmetric, we further have that $\mathcal{I}_1 \supseteq \mathcal{I}_2 \supseteq \mathcal{I}_3$.
    
    Alternatively, consider the weighted-by-group-size DID estimator, inference based on a parametric heteroskedasticity model given by group-size variation ($\mathbb{V}[W_s] = A + B/M_s$) and the resulting confidence interval obtained from inverting test procedures 1 to 4. We then have that $\mathcal{I}_1 \supseteq \mathcal{I}_3 \supseteq \mathcal{I}_4$ and $\operatorname{length}(\mathcal{I}_1)  \geq \max\{\operatorname{length}(\mathcal{I}_2), \operatorname{length}(\mathcal{I}_3), \operatorname{length}(\mathcal{I}_4) \}$. In addition, if the empirical distribution of the error term is symmetric, we further have that $\mathcal{I}_1 \supseteq \mathcal{I}_2 \supseteq \mathcal{I}_3 \supseteq \mathcal{I}_4$.
\end{proposition}
\begin{proof}
    See Appendix \ref{proof_inclusion}.
\end{proof}

\section{Assessing the plausibility of Assumption \ref{As_W}.(i)}
\label{app_plausible_het}
Observe that an implication of Assumption \ref{As_W}.(i) is that there exist a quantile function $h: (0,1) \mapsto \mathbb{R}$, such that
\begin{equation}
	\label{eq_imp}
Q_{s}(u) = h(u), \quad \forall s \in \mathcal{I}_0, u \in (0,1) \, ,
\end{equation}
where $Q_{s}$ is the quantile function of $|W_s|/h(Z_s;\delta)$. This suggests the following procedure to assess the plausibility of a proposed heteroskedasticity correction in accounting for distribution differences across units. Using the rescaled residuals in the control group, we may perform a quantile regression of $\hat \xi_s$ on a transformation of $Z_s$, for different quantiles $u$. Given \eqref{eq_imp}, we would not expect estimated quantiles to vary as a function of the $Z_s$, for any of the $u$.

We apply the proposed test to the data used to construct our Monte Carlo exercise in Section \ref{simulations}. Specifically, we consider $u \in \{0.9, 0.95\}$, and perform quantile regression of the $\hat \xi_s$ on quintile indicators of the $M_s$. We also consider a similar procedure on the standardized residuals $\hat W_s/ \operatorname{sd}(\hat{W}_s)$. In this case, if variation in $M_s$ has a meaningful effect on the variance of $W_s$, then we should expect this quantile function not to be constant. 

\begin{figure}[H] 
	
	\begin{center}
		\caption{{\bf Quantiles of residuals as a function of PUMA size}}  \label{Figure_Quantiles}
		
		\begin{tabular}{cc} 
			
			A: $u=0.9$ &   B: $u=0.95$\\
			
			\includegraphics[scale=0.8]{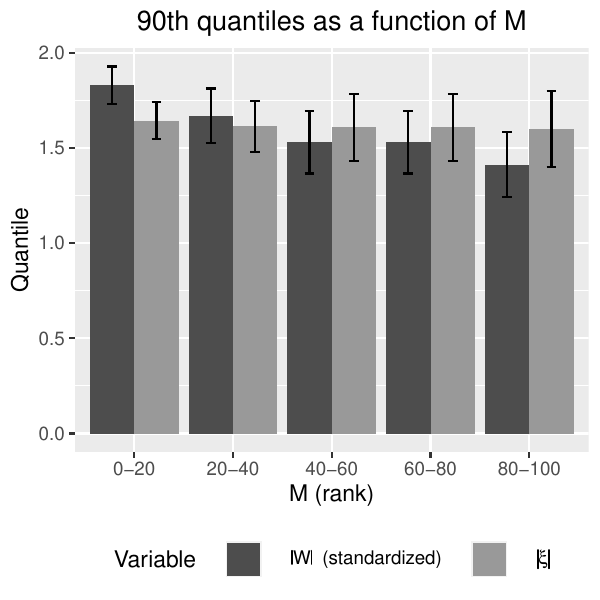} & \includegraphics[scale=0.8]{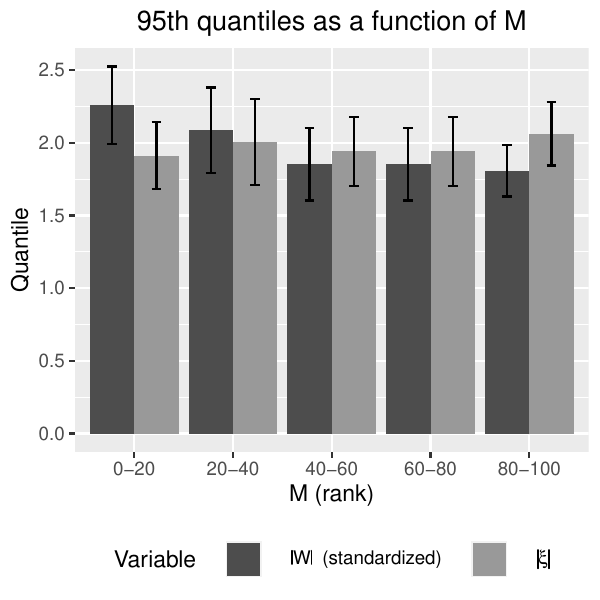} 
			
		\end{tabular}

	\end{center}

\end{figure}

Figure \ref{Figure_Quantiles} presents the estimated quantile regressions, along with 95\% confidence intervals computed under the assumption of independent observations. Clearly, adopting the parametric form of heteroskedasticity offers improvements over using the $\hat W_s$, as regards the plausibility of Assumption \ref{As_W}.(i). Indeed, the quantiles of $\hat{\xi}$ vary less than those of $\hat{W}$, especially at $u=0.9$. To further present evidence of this improvement, Table \ref{tab_pvalues} reports, for each $u$, the p-values of the null that there is no variation of the conditional quantile function across quintiles of $M$. We report p-values under the assumption of independence, as well as p-values clustered at the state level, which we compute using the bootstrap procedure in \cite{Hagemann2017} as implemented in R package \texttt{quantreg} \citep{Koenker2022}. We see that adopting the proposed parametric correction leads to improved credibility of Assumption \ref{As_W}.(i).

\begin{table}[H]
\begin{center}
	\caption{p-values: null of constant quantile function}
	\label{tab_pvalues}
	\begin{tabular}{ccccc}
		& \multicolumn{2}{c}{u = 0.9} & \multicolumn{2}{c}{u = 0.95}  \\
		\cmidrule(lr){2-3} \cmidrule(lr){4-5} 
		& independence & clustered at state level 	& independence & clustered at state level \\
		\hline
		$\hat W$& 	 0.001     &       0.003 &	   0.055       &         0.135 \\
	$\hat \xi$&	0.986      &        0.953 &    0.907        &        0.936 \\
		\hline
	\end{tabular}
\end{center}
\end{table} 

We conducted a similar exercise for the data used in our empirical illustration of Section \ref{illustration} and also found suggestive evidence of the appropriateness of the heteroskedasticity correction in this setting. Specifically, at the $u = 0.95$ quantile, the p-value clustered at state level of the null that the quantile function does not vary across quintiles of county size is $0$ for the unrescaled residuals $\hat W_s$ of both outcomes considered; and $0.20$ and $0.26$ for the rescaled residuals of, respectively, all-cause and healthcare-amenable mortality.

\section{On Assumptions \ref{As_reg} and \ref{As_reg_cons2} }

\label{Appendix: W tilde}

\subsection{Joint distributions of $(\xi_i)_{i \in \mathcal{I}_1}$ that satisfy Assumptions \ref{As_reg} and \ref{As_reg_cons2}}
\label{app_joint_list}

The following lemma shows that Assumption \ref{As_reg} is valid for a broader set of distributions than the multivariate Gaussian case discussed in the main text.

\begin{lemma}
\label{Lemma_elliptical}
Suppose that Assumption \ref{As_W} is satisfied, and that $(\xi_{i})_{i \in \mathcal{I}_1} \overset{d}{=} A \boldsymbol{Z}$, where $A$ is an $N_1 \times L$ matrix, and $\boldsymbol{Z}$ is a zero-mean random vector taking values on $\mathbb{R}^{L}$ such that $\boldsymbol{Z}\overset{d}{=} B \boldsymbol{Z} $ for every $L \times L$ orthogonal matrix $B$. Then Assumption \ref{As_reg} is satisfied for any $\tau \in (0,1/2)$.
\end{lemma}
\begin{proof}
See Appendix \ref{proof_elliptical}.
\end{proof}

Lemma \ref{Lemma_elliptical} extends the multivariate normal case to the general setting where the joint distribution of the $\xi_i$ is given by a linear transformation of a vector of \textit{spherical} random variables, i.e. a linear transformation of random variables whose joint distribution is invariant to orthogonal transformations. Whereas a zero-mean multivariate normal distribution is known to satisfy this property, Lemma \ref{Lemma_elliptical} also covers other types of distributions, as in the case where the joint distribution of the $\xi_i$ is given by linear combinations of iid random variables with a t-distribution. While  Lemma \ref{Lemma_elliptical} can be deduced from known results in the literature (e.g. \cite{Embrechts_McNeil_Straumann_2002}), for completeness we provide a direct proof in Appendix \ref{proof_elliptical}.

The result in Lemma \ref{Lemma_elliptical} allows for a more general class of distributions than those given by the multivariate normal. However, it should be noted that the assumption that the distribution of the $(\xi_i)_{i \in \mathcal{I}_1}$ can be expressed as a linear transformation of a spherical random vector -- sometimes referred to by ellipticity in the literature -- imposes restrictions on the dependence structure among the $\xi_i$. Indeed, one of its implications is that the conditional expectation of a $\xi_i$ given $\xi_j$, $j \neq i$, is linear \citep[page 188]{Embrechts_McNeil_Straumann_2002}, which may be implausible, for example, in settings where $\xi_i$ is the result of production decisions of agents in each treated region $i$, one of the factors of production is mobile across regions, and the production process exhibits marginal decreasing returns. 

It should be noted, however, that Assumption \ref{As_reg} may also be satisfied in settings where the dependence among the $\xi_i$ is nonlinear. Indeed, note that, since the distribution $F_\xi$ has finite first moment by Assumption \ref{As_W}, there exists  $\alpha \geq 1$ and $C>0$ such that, $$\limsup_{x \to \infty} \frac{\mathbb{P}[|\xi|>x]}{x^{-\alpha}} \leq C\, ,$$

In addition, if there also exists a $c>0$ such that:
$$c \leq \liminf_{x \to \infty} \frac{\mathbb{P}[|\xi|>x]}{x^{-\alpha}} \, ,$$
we say that the distribution of $\xi$ admits a moderate power-law tail. This assumption is consistent with the distribution of $\xi$ exhibiting heavy-but-not-too-heavy tail behavior, in that tail probabilities decay polynomially fast, but not exponentially fast (as in a Gaussian distribution); yet the decay is sufficiently fast so that $\xi$ has finite first moment. Under this tail restriction, \citet{Ibragimov2017} review several sufficient conditions ensuring the existence of a threshold $\bar{\tau}$ sufficiently close to zero such that Assumption \ref{As_reg} holds for every $\tau \leq \bar{\tau}$. For example, when $N_1=2$, they show this will always be true, i.e. for sufficiently small significance level, Assumption \ref{As_W} will be satisfied. Moreover, for $N_1 > 2$, Assumption \ref{As_W} holds for sufficiently small $\tau$ whenever the dependence structure among the $\xi$ is captured by either power-type or Archimedean \textit{copulae}. 

\begin{remark}
\label{remark_el}
    A simple inspection of the proof of Lemma \ref{Lemma_elliptical} in Appendix \ref{app_lemmata} reveals that, if $\boldsymbol{\xi}$ follows an elliptical distribution, then, for any $\boldsymbol{s}\in \mathbb{R}^{N_1}$, $\frac{s'\boldsymbol{\xi}}{\sqrt{\mathbb{V}[s'\boldsymbol{\xi}]}} \overset{d}{=} \frac{\xi_1}{\sqrt{\mathbb{V}[\xi_1]}}$, where $\xi_1$ is shown to be symmetric about zero. This implies that linear combinations $s'\boldsymbol{\xi}$ of $\boldsymbol{s}\in \mathbb{R}^{N_1}$ follow a scale family with base quantile function $Q_{\xi}$ that satisfies $Q_{\xi}(1-u)\geq 0 \geq Q_{\xi}(u)$ for any $u \in (0,1/2)$. Consequently, if the variance bound $\mathbb{V}(\mathcal{W}) \leq A + B / {M_{\mbox{\tiny T} }}$ is satisfied and $\boldsymbol{\xi}$ follows a symmetric distribution, it must be that Assumption \ref{As_reg_cons2} holds for any $\tau \in (0,1/2)$. This shows that, under the variance bound in $\mathbb{V}(\mathcal{W}) \leq A + B / {M_{\mbox{\tiny T} }}$, the asymptotic validity of Method 4 in the main text is justified under an elliptical joint distribution for the cluster average errors.
\end{remark}

\subsection{Assessing the plausibility of Assumptions \ref{As_reg} and \ref{As_reg_cons2} in parametric classes of \textit{copulae}}

\label{App_copulas}

Notice that the results surveyed in the previous section -- and, more generally, those available in the literature --, do not cover not the family of Gaussian \textit{copulae}, a popular class for modeling dependence structures in applied research.\footnote{An exception in the literature is \citet{Asmussen2008}, who shows that, if the tails of $|\xi|$ behave as the right-tail of a log-normal distribution, then Assumption \ref{As_reg} will be satisfied for sufficiently small $\tau$.} To cover this case, this Appendix provides an approach to assess the plausibility of Assumptions \ref{As_reg} and \ref{As_reg_cons2} in particular applications under the assumption that the true dependence structure belongs to a posited family of \textit{copulae} -- with Gaussian \textit{copulae} being the leading case in mind. 
 
To introduce our approach, we begin by observing that, under either Assumption \ref{As_reg} or \ref{As_reg_cons2}, the marginal distribution of the $W_s$, $s \in \mathcal{I}_1$, is identified. Indeed, since the normalized distribution of residuals in the control group is a consistent estimator of $F_{\xi}$ (Proposition \ref{Prop_N1_FP}), and given that $W_s = h(Z_s; \delta) \xi_s$ and the parametric estimator of the heteroskedasticity is consistent, it follows that the marginal distribution of the $W_s$ is consistently estimated as $\hat F_{W_s}(x) = \hat F_{\xi}(x/h(Z_s;\hat \delta ))$, $x \in \mathbb{R}$. Using this fact, we propose a method to assess the plausibility of Assumptions \ref{As_reg} or \ref{As_reg_cons2} under a wide range of dependence structures for the $W_s$. 

The proposed method proceeds as follows. Let $\hat q_{\xi}$ denote the empirical quantile functions of the normalized residuals in the control group. Let $H$ be a $N_1$-dimensional \textit{copula}, a cdf on $[0,1]^{N_1}$ with uniform marginals. For $b=1,\ldots, B$, we may construct samples:

$$(W^b_1, \ldots, W^b_{N_1}) = \left(h(Z_1;\hat \delta )\hat{q}_\xi(U_1^b), \ldots, h(Z_{N_1};\hat \delta )\hat{q}_\xi (U_{N_1}^b)\right) \, , \quad (U_1^b,\ldots, U_{N_1}^b) \sim H \, .$$

Observe that, as $N_0 \to \infty$, observations in each of these samples have  marginal distributions equal to the marginals of the $\{W_s\}_{s \in \mathcal{I}_1}$, with dependence structure dictated by $H$. We can then use these $B$ samples to verify the validity of Assumptions \ref{As_reg} and \ref{As_reg_cons2} under dependence structure $H$. For example, we will say dependence structure $H$ satisfies \ref{As_reg} if:
{\footnotesize
$$  \tilde{p}_H = \frac{1}{B}\sum_{b=1}^B \mathbf{1}\left\{ \left\{\frac{1}{N_1}\sum_{i=1}^{N_1} W^b_i >  \left(\frac{1}{N_1}\sum_{i=1}^{N_1} h(Z_s, \delta) \right)\hat{q}_{\xi}(1-\tau/2) \right\}\cap \left\{\frac{1}{N_1}\sum_{i=1}^{N_1} W^b_i < \left(\frac{1}{N_1}\sum_{i=1}^{N_1} h(Z_s, \delta) \right)\hat{q}_{\xi}(\tau/2) \right\}\right\} \leq \tau \, ,$$}
and similarly for Assumption \ref{As_reg_cons2}. By varying $H$, we may then assess the validity of the assumptions under different dependence structures.

For the procedure to convey meaningful information, we must find a method to vary $H$ in a principled manner. One alternative is to sample $H$ from a nonparametric class of copulas. When $N_1=2$, \cite{Guillotte2012} provides a sampler from a Jeffrey's prior on a nonparametric class of bivariate copulas. For $N_1 >2$, one may use one of the priors available in nonparametric Bayesian Statistics \citep[e.g.][]{ghosal2017fundamentals} to sample from a nonparametric class of multivariate cdfs, which can then be normalized to have uniform marginals. However, it should be noted that the computational cost of these approaches is increasing in $N_1$ and can be quite large even for few treated units.

Another alternative is to consider parametric classes of copulas. This is the approach we undertake when verifying the plausibility of Assumptions \ref{As_reg} and \ref{As_reg_cons2} on the data used to construct our simulations in Section \ref{simulations}. Specifically, we vary $H$ over the class of Gaussian copulas, which are able in our data to produce non-monotone conditional expectations between the treated units' $\xi_s$. This class is given as follows. For a given positive semidefinite $N_1 \times N_1$ matrix $\Sigma$ with trace equal to $N_1$, the draws $(U_i^b)_{i=1}^{N_1}$ are given by:

\begin{equation*}
	\begin{aligned}
		(U_1^b, \ldots, U_{N_1}^b) = \left(\Phi\left(\frac{Z_1^b}{\Sigma_{11}}\right),\ldots \Phi\left(\frac{Z_1^b}{\Sigma_{N_1 N_1}}\right)\right) \, ,  \quad
		(Z_1,\ldots, Z_{N_1}) \sim \mathcal{N}(0, \Sigma) \, .
	\end{aligned} 
\end{equation*}

We can then span this class of copulas by sampling from the class of $N_1 \times N_1$ positive semidefinite matrices with trace equal to $N_1$. In \texttt{R}, we may do this by first sampling an $N_1 \times N_1$ orthogonal matrix using the \texttt{rorth} function available in package \texttt{ICtest}, which generates draws with respect to the Haar measure on the space of orthogonal $N_1 \times N_1$  matrices \citep{Nordhausen2022}. We then sample a $N_1$-dimensional vector uniformly from the $N_1-1$ simplex, and multiply the draw by $N_1$. Finally, we can use the spectral decomposition to construct the resulting matrix $\Sigma$. Using this procedure, and for different values of $N_1$ and the chosen treated units in Section \ref{simulations}, we consider $100,000$ Gaussian copulas drawn at random. We verify that all such dependence structures satisfy Assumption  \ref{As_reg}. Once we discard copulas that do not satisfy the variance bound $\mathbb{V}(\mathcal{W}) \leq A + B / {M_{\mbox{\tiny T} }}$ (i.e. we discard those structures where within-state correlation between individuals is smaller than between-state correlation), we also verify that Assumption \ref{As_reg_cons2} is always satisfied.

We conduct a similar exercise on the dataset used in our empirical illustration, and verify that Assumptions \ref{As_reg} and \ref{As_reg_cons2} are always satisfied for all simulated copulas and both outcomes considered.

If in a given empirical application this simulation procedure detects violations of the required assumptions, a simple correction is available. Let $\mathcal{H}$ be the set of copulas for which the assumptions were verified. We can then consider the corrected critical value $\tilde{\tau} := \sup_{h \in \mathcal{H}} \tilde{p}_H$ when conducting inference. The resulting procedure will be asymptotically valid whenever the true dependence structure is less conservative than the most conservative element in $\mathcal{H}$.

\subsection{A simple empirical check}
\label{Appendix_simple_check}

Instead of computing worst-case rejection rates over a class of \textit{copulae}, in this section we propose yet another exercise to assess the plausibility of Assumption \ref{As_reg} in the  ACS dataset used to construct the Monte Carlo exercise in the MC simulations presented in Section  \ref{simulations}. As we discuss in detail in Appendix \ref{Appendix_MC}, to conduct the MC exercise, we calculate the residuals $\widehat W_s$ at the PUMA level with the ACS data, and then estimate a model for the spatial correlation across PUMAs. We may use this estimated model to assess the plausibility of Assumption   \ref{As_reg} under some ``extreme'' dependence structures. Specifically, we first calculate the $\tau/2$ and $1-\tau/2$ quantiles of the distribution of $\{ \widehat W_s \}$, $c_{\tau/2}$ and $c_{1-\tau/2}$. Then, we use the estimated correlation matrix to identify for each PUMA the other PUMA that is most correlated with it, and  calculate the proportion of pairs in which the average of these two PUMAs is smaller than $c_{\tau/2}$ or greater than $c_{1-\tau/2}$. For all $\tau \in \{ 0.01,0.02,\hdots,0.99\}$, we find that this proportion is smaller than $\tau$. We reach the same conclusion if we consider the average of two random PUMAs. Therefore, Assumption \ref{As_reg} seems reasonable in the application we consider for our MC simulations, even for these two ``extreme'' dependence structures considered herein.

\subsection{An example where Assumption \ref{As_reg} fails}
\label{Appendix_reg}

 One case in which Assumptions \ref{As_reg} may not be valid for some values of $\tau$ is when the distribution of $W_s$ is bimodal, with the two peaks very far apart. To illustrate that, consider the case in which $W_s$ is $N(-4.167,1)$ with probability 0.96 and $N(100,1)$ with probability 0.04. Therefore, we have that $\mathbb{E}[W_s]=0$.  If we consider $\tau = 0.1$, then $c_{0.05} \approx -5.792$ and $c_{0.95} \approx -1.857$. If we use these critical values to calculate $\tilde p$ when we have $N_1=5$, then we have $\tilde p \approx 0.18$. This happens because the probability that at least one of the observations come from the distribution with mean 100 is approximately $18\%$ and, conditional on that, the probability that $\widetilde W >c_{0.95}$ is close to one. 

We stress, however, that this is a distribution for the errors that we engineered to provide an example of a distribution in which Assumption \ref{As_reg} would not hold, and that this is not a kind of distribution for the errors we should expect in common  applications.  To construct this example, we need not only a bimodal distribution, but also  the distance between that peaks to be very large. If instead we consider a setting in which  $W_s$ is $N(-0.042,1)$ with probability 0.96 and $N(1,1)$ with probability 0.04 (so, again, $\mathbb{E}[W_s]=0$), then we would have $\tilde p \approx 0$. If we had instead that the mean of the second peak is 10, then we would have $\tilde p = 0.066<0.1 = \tau$. Overall, this example highlights that we need very extreme and unrealistic distributions for the errors so that Assumption \ref{As_reg} fails. Note also that in empirical applications it would be possible to observe the marginal distribution of $\widehat W_s$, and check whether it has this kind a shape that could invalidate Assumption \ref{As_reg}. If we  find evidence that this marginal distribution does not have peaks or large mass on the tails, then we should be more comfortable with Assumption \ref{As_reg}.

\subsection{Microfounding Assumption \ref{As_reg_cons2}}
\label{app_micro}
Consider a simple setting where the pre-post difference of unobservables in state $s$ is given by:

$$W_s = \frac{1}{M_s}\sum_{i=1}^{M_s} \epsilon_{i,s}\, ,$$
where $\epsilon_{i,s}$ denotes the unobservables corresponding to the  pre-post evolution of outcomes of the $i$-th individual in cluster $s$. We assume that the $\epsilon_{i,s}$ have homogeneous variance $\sigma^2_\epsilon$ across units and states, and that the correlation between the unobservables of two distinct units in the same cluster is homogeneous and given by $\rho$, whereas the correlation of any two individual unobservables in clusters $s\neq s'$ is given by $\psi(s,s')$. Finally, we assume that the vector stacking the $(\epsilon_{i,s})_{i=1}^{M_s}$ across $s \in \mathcal{I}_1$ has an elliptical distribution. In this case, as discussed in Remark \ref{remark_el}, it suffices to verify the variance bound $\mathbb{V}[{W}]\leq A + \frac{B}{M_{\tiny T}}$ to ensure the validity of Assumption $\ref{As_reg_cons2}$.

Observe that, under our assumptions, $A = \rho \sigma^2_{\epsilon}$ and $B = (1-\rho)\sigma^2_\epsilon$. Moreover, we obtain that the covariance between $W_{s}$ and $W_{s'}$, $s \neq s'$, is given by:

$$\operatorname{cov}(W_s, W_{s'}) = \psi(s,s')\sigma^2_{\epsilon}\, . $$

Consequently, we have that, by defining $\omega_s =M_s/M_T$:

\begin{equation}
    \begin{aligned}
      \mathbb{V}[{W}]= \sum_{s\in \mathcal{I}_1} \omega_s^2 [\rho \sigma^2_\epsilon + (1-\rho)\sigma^2_\epsilon/M_s ] + \sum_{(s,s') \in \mathcal{I}_1^2: s \neq s'} \omega_s\omega_{s'}\psi(s,s')\sigma^2_\epsilon =  \\ \left[ \sum_{s\in \mathcal{I}_1} \omega_{s}\left(\omega_s \rho+  \sum_{s'\in \mathcal{I}_1: s' \neq s} \omega_{s'}\psi(s,s')\right) \right]\sigma^2_\epsilon   + \frac{B}{M_T}\, ,
    \end{aligned}
\end{equation}
consequently, if $\psi(s,s') \leq \rho$ for every $s,s' \in \mathcal{I}_1$, $s \neq s'$, meaning that the correlation of two units in the same cluster is always larger than the correlation in different clusters, we have that:

$$\left[ \sum_{s\in \mathcal{I}_1} \omega_{s}\left(\omega_s \rho+  \sum_{s'\in \mathcal{I}_1: s' \neq s} \omega_{s'}\psi(s,s')\right) \right]\sigma^2_\epsilon \leq A\, ,$$
proving that the variance bound is satisfied.

\section{Method 4 with variation in observations per cell across time} \label{Appendix_cons2}

If the number of observations per cell, $M_{st}$, varies with $t$, we recommend first aggregating the data at the unit $\times$ time level, i.e. we construct a unit-time level dataset where each observation corresponds to $\tilde{Y}_{s,t} = \frac{1}{M_{s,t}}\sum_{i=1}^{M_{s,t}}Y_{i,s,t}$. Then, in order to provide bounds on the variance of the estimator in the presence of spatial correlation, we recommend computing a weighted TWFE/DID estimator where each pair $(s,t)$ receives weight equal to $ \min_{t \in \mathcal{T}_0 \cup \mathcal{T}_1}M_{s,t}$. In this case, we have that, as $N_0 \to \infty$, the weighted DID estimator $\hat \alpha_w$ is consistent to:

$$ \hat \alpha_w \overset{p}{\to} \alpha_\omega  + \sum_{s \in \mathcal{I}_1} \omega_s W_s \, , $$
where $\alpha_\omega = \sum_{s \in \mathcal{I}_1}\omega_s \frac{1}{T_1}\sum_{t \in \mathcal{T}_1}\frac{1}{M_{s,t}}\sum_{i=1}^{M_{s,t}} \alpha_{i,s,t}$ is a weighted average treatment effect on the treated parameter, with weights $\omega_s = \frac{\min_{t \in \mathcal{T}_0 \cup \mathcal{T}_1}M_{s,t}}{\sum_{s' \in \mathcal{I}_1} \min_{t \in \mathcal{T}_0 \cup \mathcal{T}_1}M_{s',t}}$; and $W_s = \frac{1}{T_1 }\sum_{t \in \mathcal{T}_1} \frac{1}{M_{s,t}}\sum_{i=1}^{M_{s,t}} \epsilon_{i,s,t} - \frac{1}{T_0 }\sum_{t \in \mathcal{T}_0} \frac{1}{M_{s,t}}\sum_{i=1}^{M_{s,t}} \epsilon_{i,s,t}$, with $\epsilon_{i,s,t}$ denoting the time-varying individual unobservable driving variation in the untreated potential outcome.

The next step in extending Method 4 consists in specifying a model for the variance of $W_s$. We consider the variance of $W_s$ under the assumption that individual-level shocks $\epsilon_{i,s,t}$ have homogeneous variance $\sigma^2_{\epsilon}$, common correlation $\rho$ with other individual shocks in the same unit and time period, and correlation $\delta(t,t')$ with individual shocks in the same unit, but time period $t'$.\footnote{The correlation may be seem as capturing common shocks that affect all units in each cross-section and are serially correlated, or, in rotating panels, the fact that a fraction of the units may appear repeatedly over time.} In this case, we are able to show that:
\begin{equation*}
    \begin{aligned}
        \mathbb{V}[W_s] = \sigma^2_\epsilon(1-\rho) \left[\frac{1}{T_1^2}\sum_{t \in \mathcal{T}_1}\frac{1}{M_{s,t}} +  \frac{1}{T_0^2}\sum_{t \in \mathcal{T}_0}\frac{1}{M_{s,t}}\right]  + \\ \sigma^2_\epsilon \left[\rho\left(\frac{1}{T_1}+\frac{1}{T_0}\right) + \frac{1}{T_1^2}\sum_{(t,t')\in \mathcal{I}_1^2: t \neq t'}\delta(t,t') + \frac{1}{T_1^2}\sum_{(t,t')\in \mathcal{I}_0^2: t \neq t'}\delta(t,t') - 2  \frac{1}{T_1 T_0} \sum_{(t,t')\in \mathcal{I}_0\times \mathcal{I}_1} \delta(t,t')\right]  = \\{B}_*\left[\frac{\frac{1}{T_1^2}\sum_{t \in \mathcal{T}_1}\frac{1}{M_{s,t}} + \frac{1}{T_0^2}\sum_{t \in \mathcal{T}_0}\frac{1}{M_{s,t}}}{\frac{1}{T_0}+\frac{1}{T_1}}\right] + {A}_* \, ,
    \end{aligned}
\end{equation*}
which suggests considering a parametric heteroskedasticity model where variation stems from the term $\left(\frac{1}{T_1^2}\sum_{t \in \mathcal{T}_1}\frac{1}{M_{s,t}} + \frac{1}{T_0^2}\sum_{t \in \mathcal{T}_0}\frac{1}{M_{s,t}}\right)/(1/T_0 + 1/T_1)$. The term ${A}_*$ in this parametrization captures within-cluster correlation in unobservables.

Having estimated the above parametrization, we next observe that, under the assumption that $\operatorname{cov}(W_s, W_{s'}) \leq A_*$ for $s \neq s'$, i.e. that spatial correlation between units in different clusters is smaller than that in units in the same cluster, we have that:

\begin{equation}
\label{var_inequality_weights}
    \begin{aligned}
       \mathbb{V}\left[ \sum_{s \in \mathcal{I}_1} \omega_s W_s  \right] = 
 \sum_{s \in \mathcal{I}_1} \omega_s^2 \mathbb{V}[W_s] + \sum_{s \in \mathcal{I}_1^2: s\neq s'}\omega_s \omega_{s'} \mathbb{C}(W_s,W_{s'}) \leq \\ A_{*} + \sum_{s \in \mathcal{I}_1}\omega_s^2 B_* \left[\frac{\frac{1}{T_1^2}\sum_{t \in \mathcal{T}_1}\frac{1}{M_{s,t}} +  \frac{1}{T_0^2}\sum_{t \in \mathcal{T}_0}\frac{1}{M_{s,t}}}{\frac{1}{T_0}+\frac{1}{T_1}}\right]   \leq A_* + B_*\frac{1}{\sum_{s' \in \mathcal{I}_1}\min_{t \in \mathcal{I}_0\cup \mathcal{I}_1}\mathcal{M}_{s',t}}\, .
    \end{aligned}
\end{equation}

This suggests a procedure for testing the null $H_0: \alpha_\omega = \alpha_0$ against the one sided alternative where one rejects the null if $\hat{\alpha}_w - \alpha_0$ exceeds the estimated tail quantiles of $\sqrt{A_* + B_*\frac{1}{\sum_{s' \in \mathcal{I}_1}\min_{t \in \mathcal{I}_0\cup \mathcal{I}_1}\mathcal{M}_{s',t}}} \xi$, where $Q_\xi(\cdot)$ denotes the quantile function of the standardized error $\xi = W/\sqrt{\mathbb{V}[W]}$. The asymptotic conservativeness of this procedure is justified under a suitable adaption of Assumption \ref{As_reg_cons2}. Specifically, one replaces Assumption \ref{As_reg_cons2} with the requirement that the probability   that $\sum_{s' \in \mathcal{I}_1}\omega_{s'} W_{s'}$ attains extreme values is smaller than the probability that $\sqrt{A_* + B_*\frac{1}{\sum_{s' \in \mathcal{I}_1}\min_{t \in \mathcal{I}_0\cup \mathcal{I}_1}\mathcal{M}_{s',t}}}\xi$ would attain such values.

\begin{remark}[Alternative weights]
    An inspection of the variance inequality \eqref{var_inequality_weights} that motivates our testing procedure reveals that one could alternatively rely on a weighted DID estimator with unit-time weights proportional to:
    $$\frac{1}{\frac{1}{T_1^2}\sum_{t \in \mathcal{T}_1}\frac{1}{M_{s,t}} + \frac{1}{T_0^2}\sum_{t \in \mathcal{T}_0}\frac{1}{M_{s,t}}}\, .$$

    In this case, one rejects the null if $\hat{\alpha}_w - \alpha_0$ exceeds the tail quantiles of $$\sqrt{A_* + B_*\frac{[T_0^{-1}+T_1^{-1}]^{-1}}{\sum_{s'\in \mathcal{I}_1}\left(\frac{1}{T_1^2}\sum_{t \in \mathcal{T}_1}\frac{1}{M_{s',t}} + \frac{1}{T_0^2}\sum_{t \in \mathcal{T}_0}\frac{1}{M_{s',t}}\right)^{-1}}} \xi  \, .$$

    In principle, this procedure should be ``less conservative'' than the one that uses the weights given by $\min_{t}M_{s,t}$, as it relies on a tighter variance bound. However, we believe this comes at a cost in terms of interpreting the weighted average treatment effect being estimated.
\end{remark}

\section{Conservative tests with variation in treatment timing}
\label{Appendix_staggered}

Constructing a conservative test requires more care when treated units start treatment at different periods. For example, consider that unit 1 starts treatment after $t_1$, while unit 2 starts treatment after $t_2$. In this case, the asymptotic distribution of $\widehat \alpha$ would depend on the linear combinations of the errors $W_1(1) = \frac{1}{T-t_1} \sum_{t=t_1+1}^T \eta_{1t}-\frac{1}{t_1} \sum_{t=1}^{t_1} \eta_{1t}$, and $W_2(2) = \frac{1}{T-t_2} \sum_{t=t_2+1}^T \eta_{2t}-\frac{1}{t_2} \sum_{t=1}^{t_2} \eta_{2t}$. We can still consistently estimate the marginal distributions of $W_1(1)$ and $W_2(2)$ by considering the appropriate linear combination of the residuals from the control units (see FP and \cite{alvarez2023extensions} for discussions), and the Methods in Section \ref{bounds_arbitrary} remain valid under suitable adaptation of the formulae. Similarly, the method in Section \ref{Bounds_positive} remain valid if one assumes the positive-dependence condition between $|W_1(1)|$ and $|W_1(2)|$; and, the method in Section \ref{Sec_cons1} can be adopted by considering a comonotone copula for $W_1(1)$ and $W_1(2)$ as the worst-case scenario for spatial correlation. Notice, however, that, differently from the uniform treatment setting, such choice of comonotone copula does not necessarily correspond to a scenario where  $\mbox{corr}(\eta_{1t},\eta_{2t})=1$. To see that, suppose  there are 3 time periods, with $t_1 = 1$ and $t_2 = 2$. In this case, $\widehat \alpha - \alpha \buildrel p \over \rightarrow 0.5(0.5 \eta_{13}+0.5 \eta_{12} - \eta_{11}) + 0.5(\eta_{23}-0.5 \eta_{22} - 0.5 \eta_{21})$. Therefore, assuming $\mbox{corr}(\eta_{1t},\eta_{2t})=1$ will lead to a \emph{lower} variance relative to the case with no spatial dependence if $\mathbb{V}(\eta_{i2})$ is substantially larger than the variance at the two other periods. Therefore the interpretation of the comonotone assumption should be made directly in terms of the $W_1(1)$ and $W_1(2)$ in this case.

As for implementation of the method in Section \ref{Sec_cons2}, we recommend a two-step approach. In a first-step, we partition the set of treated clusters $\mathcal{I}_1$ into cohorts, with the property that any two clusters in the same cohort start treatment at the same period. For clusters in the same cohort, the relevant pre-post difference in errors will involve the same pre- and post-treatment window (see \cite{alvarez2023extensions} for details). One can therefore apply Method 4 to bound the quantiles of linear combinations of errors of clusters in the same cohort, under the assumption that the tail quantiles of this combination are bounded by the quantiles of the distribution of a hypothetical single cluster with total number of individual-level observations equal to the total number of individual-level observations in the cohort. It then remains to combine bounds from different cohorts to produce a single bound for the quantiles of the linear combination of errors across all treated units. In a second step, we therefore suggest that researchers combine bounds for different adoption dates using one of the methods in Sections \ref{bounds_arbitrary}, \ref{Bounds_positive} or \ref{Sec_cons1}.

\section{Details on the MC simulation} \label{Appendix_MC}

We construct our MC simulations based on the  American Community Survey (ACS). We restrict our sample to individuals on the worforce, and consider log wages as the outcome variable. With data from 2014 to 2017, we estimate the residuals $\widehat W_s$ at the PUMA level, considering a setting in which treatment starts after 2015. More specifically, we compute the residuals from a PUMA $\times$ year regression. We then calculate  $\widehat W_s = (\hat e_{s,2017} + \hat e_{s,2016})/2 - (\hat e_{s,2015} + \hat e_{s,2014})/2$. 

Given $\widehat W_s$ for all PUMAs, we estimate a model for the spatial correlation in which the covariance between two PUMAs may depend on the similarity between their industry compositions, on whether they belong to the same state, and on their worforce sizes. More specifically, letting $M_s$ denote the (average) workforce size at PUMA $s$, we assume that $(W_1,...,W_N)$ is multivariate normally distributed with mean zero and variance/covariance matrix given by:

\begin{eqnarray}
	\label{eq_model_cov}
	\mbox{cov}(W_s,W_h) = \begin{cases}  \sigma_P^2 + \frac{\sigma_\epsilon^2}{M_s} + \sigma_S^2  + \sigma^2_I & \mbox{ if } s=h   \\
		\sigma_S^2 + \sigma^2_I \mu_s' \mu_h    & \mbox{ if } s \neq h \mbox{ and } \text{state}_s = \text{state}_h   \\
		\mu_s' \mu_h    & \mbox{ if } \text{state}_s \neq \text{state}_h   			        
	\end{cases} ,
\end{eqnarray}
where $\mu_s$ is a $12\times 1$ vector of PUMA industry composition measures normalized so $\mu_s'\mu_s=1$. These industry measures are obtained by performing principal component analysis on the PUMA sector shares computed according to the Census 1990 industry classification. Parametrization \eqref{eq_model_cov} can be motivated by the following factor model for the $W_s$ \citep{FermanJAE}:

$$W_s = \omega_s + \frac{1}{M_s} \sum_{i =1}^{M_s} \epsilon_{i,s} + \gamma_{\text{State}_s} + \mu_s' \psi \, ,$$
where $\omega_s$ is a PUMA-level shock, $\epsilon_{i,s}$ are individual-level shocks, $\gamma_{\text{State}_s} $ is a state shock, and $\psi$ is a vector of industry shocks with common variance. If shocks are assumed to be uncorelated, then this factor model leads to parametrization \eqref{eq_model_cov}.

We estimate \eqref{eq_model_cov} by nonnegative least-squares. We find that:
$$(\hat \sigma_P^2,\hat \sigma_\epsilon^2, \hat  \sigma_S^2,\hat \sigma_{I}^2)= (6.627967 \times 10^{-4}, 2.400695 \times 10^2, 2.022981 \times 10^{-4},  1.870911 \times 10^{-4}).$$

For the MC simulations presented in Section \ref{simulations} we first selected a  PUMA that exhibits a strong spatial correlation with some other PUMAs to be treated.  {Then we sequentially pick $N_1-1$ PUMAs in the same state that are most similar in industry composition with previously selected PUMAs.} Given this treatment assignment, we draw random normal variables $(W_1,...,W_N)$ from the estimated DGP.  We note that all simulations are invariant to the choice of unit and time fixed effects. Moreover, we consider simulations in which the treated units are fixed, treatment effects are also fixed (and equal to zero), and the only stochastic components are the $W_s$. Therefore, these simulations are based on the exact  same framework as in the theoretical part of the paper. All TWFE regressions are weighted by workforce size.

\section{Lemmata}
\label{app_lemmata}
Below we provide two lemmas that are used throughout our proofs. Their proofs can be found in Appendix \ref{app_proofs}.
\begin{lemma}
\label{lemma_uniform_qt}
    Let $F_n$ be a sequence of random distribution functions. Suppose there exists a continuous distribution function $F$ such that, for every $x \in \mathbb{R}$ and as $n \to \infty$:

    $$F_n(x) - F(x) \overset{p}{\to} 0 \, .$$

    We then have that, for any $0 < \underline{p} < \overline{p}<1$:

    $$\sup_{p \in [\underline{p},\overline{p}]} |Q_{F_n}(p) - Q_F(p)| \overset{p}{\to} 0\, .$$

    Moreover, if $F$ has bounded (right-bounded/left-bounded) support, the above is also true for $0=\underline{p}<\overline{p}=1$ ($0=\underline{p}<\overline{p}<1$ / $0<\underline{p}<\overline{p}=1$).
\end{lemma}
\begin{proof}
    See Appendix \ref{proof_uniform_qt}.
\end{proof}
\begin{lemma}[Sufficient conditions for integrated $L^1$--consistency of the empirical quantile process]
\label{lemma_ES_CONVERGENCE}
Let $\hat Q_n$ be a sequence of random quantile functions of a sample of $n$ identically distributed scalar random variables $X_1,X_2,\ldots, X_n$, where $X_i \sim F$. Denote by $Q$ the quantile function of $F$, and suppose that, for some $\tau \in (0,1)$, the following properties hold.
\begin{enumerate}
    \item[\textnormal{(i)}] (\emph{Interior uniform consistency}) 
    For every $c\in [\tau,1)$,
    \[
    \sup_{u\in[\tau,c]} |\hat Q_n(u)-Q(u)| \xrightarrow{p} 0.
    \]
    \item[\textnormal{(ii)}] (\emph{Upper-tail integrability}) 
    \[
     \mu^+_\tau :=  \int_\tau^1 |Q(u)|\,du  < \infty.
    \]
\end{enumerate}
Then
\[
\int_{\tau}^1 \!|\hat Q_n(u)-Q(u)|\,du \;\xrightarrow{p}\; 0.
\]
\end{lemma}
\begin{proof}
  See Appendix  \ref{proof_es_conv}.
\end{proof}

\section{Proof of results in the Appendix}
\label{app_proofs}
\subsection{Proof of Proposition \ref{prop_makarov}}
\label{proof_makarov}
\begin{proof}
  It suffices to show that, as $N_0\to \infty$, $\hat U \overset{p}{\to} \bar{Q}_{\widetilde{W}}(1-\tau/2)$ and $\hat L \overset{p}{\to} \bar{Q}_{\widetilde{W}}(\tau/2)$, as the conclusion then follows by a similar argument to the proof of Corollary \ref{Cor_N1}. 
    
    We analyze consistency of $\hat U$, as the case of $\hat L$ is analogous and thus omitted. If $F_\xi$ has bounded support, the conclusion follows immediately from the continuous mapping theorem, since, by Proposition \ref{Prop_N1_FP} and Lemma \ref{lemma_uniform_qt}:

    $$\sup_{u \in [1-\tau/2,1]}|\hat{Q}_\xi(u) - Q_\xi(u)| = o_{\mathbb{P}}(1)\, .$$

    Consider now the case where the support is right-unbounded. In this case, the infimum defining the population Makarov bound is attained at some $p^* \in (1-\tau/2,1)$.   Let $s\mapsto M(s)$ denote the population objective of the optimization defining the Makarov bound, and $s \mapsto \hat{M}(s)$ its sample analog. Since $\lim_{s \to 1}M(s)=\infty$ and $Q_\xi$ is continuous on $1-\tau/2$, there exists $\tilde{p} \in (p^*,1)$ such that $3M(p^*) <  M(\tilde p)$ and $\sup_{s \in [\tilde{p},1]}|Q_\xi(1-s + (1-\tau/2)) - Q_\xi(1-\tau/2)| < \epsilon$. Now let $\epsilon:= \frac{M(\tilde{p})}{4}$, $\delta := \frac{1}{2}M(p^*) $ and notice that, combining Assumption \ref{As_W}.(iv) and the first part of Lemma \ref{lemma_uniform_qt}, we obtain that, as $N_0 \to \infty$:

\begin{equation*}
\begin{aligned}
     \mathbb{P}\Big[\left \{| \hat M(p^*) - M(p^*)| < \delta \right\} \cap \left\{|h(Z_1;\hat \delta) \hat Q_\xi(\tilde p)  - h(Z_1;\delta) Q_\xi (\tilde p)|<\epsilon \right\} \cap\\ \left\{\sup_{s \in [1-\tau/2,\tilde{p}]}|h(Z_2;\hat \delta) \hat Q_\xi(s) -h(Z_2;\delta) Q_\xi(s)|<\epsilon\right\}\Big] \to 1  
     \end{aligned}
\end{equation*}

Next, we note that, on the above event, we have:
\begin{equation*}
    \begin{aligned}    
            \inf_{s \in [\tilde p,1]}\hat{M}(s) \geq \frac{1}{2}h(Z_1;\hat \delta) \hat Q_\xi(\tilde p) + \frac{1}{2}h(Z_2;\hat \delta) \inf_{s \in [\tilde p,1]} \hat Q_\xi(1-s+(1-\tau/2)) \geq \\  
            \frac{1}{2}h(Z_1;\delta)  Q_\xi(\tilde p) + \frac{1}{2}h(Z_2; \delta) \inf_{s \in [\tilde p,1]}  Q_\xi(1-s+(1-\tau/2)) - \epsilon \geq \\
             \frac{1}{2}h(Z_1;\delta)  Q_\xi(\tilde p) + \frac{1}{2}h(Z_2; \delta) Q_\xi(1-\tilde{p}+(1-\tau/2)) - 2\epsilon  = M(\tilde{p}) - 2\epsilon >  \frac{3}{2} M(p^*) \geq \hat{M}(p^*)
    \end{aligned} \, ,
\end{equation*}
from which we conclude that:

$$\mathbb{P}\left[\inf_{s \in [1-\tau/2, 1]} \hat{M}(s) =  \inf_{s \in [1-\tau/2, \tilde{p}]} \hat{M}(s) \right]\to 1\, ,$$

Analogously, by using that $\lim_{s \to 1-\tau/2}M(s) = \infty$ and applying similar reasoning, we are able to find $\check p \in (1-\tau/2, p^*)$ such that:
\begin{equation}
  \label{eq_wpa1}
  \mathbb{P}\left[\inf_{s \in [1-\tau/2, s]} \hat{M}(s) =  \inf_{s \in [\check{p}, \tilde{p}]} \hat{M}(s) \right]\to 1\, .
\end{equation}

Given \eqref{eq_wpa1}, the consistency of $\hat U$ then follows from the continuous mapping theorem and Lemma \ref{lemma_uniform_qt}.
\end{proof}

\subsection{Proof of Lemma \ref{lemma_union bound}}
\label{proof_union_sharp}
\begin{proof}

Since the support of $F$ is right-unbounded, $Q_\xi(1)=\infty$. Consequently, the minimization program 

$$\inf_{p \in [u,1]} \frac{1}{2}\sigma[ Q_{F}(p) + Q_F(1+u - p)]\, ,$$
has an interior solution. Under our assumptions, $Q_\xi$ is differentiable on $(u,1)$. Consequently, the first-order condition holds at the minimum point $u^*$, which must satisfy:

$$\frac{1}{f(Q_F(u^*))} =  \frac{1}{f(Q_F(1+u-u^*))}\, .$$

Since $f$ is strictly monotone, it must be that $u^* = 1+u-u^* \implies u^* =  (1+u)/2$, from which follows the desired result.
\end{proof}

\subsection{Proof of Proposition \ref{prop_inclusion}}
\label{proof_inclusion}

\begin{proof}
    We consider two cases in our proof:

    \begin{enumerate}
        \item \textbf{Unweighted case.}
        The inclusion $\mathcal{I}_1 \supseteq \mathcal{I}_3$ follows from the observation that, for any $p \leq \tau/2$:
$$\hat{Q}_{\xi}(p) \lor \frac{1}{\tau/2}\int_{0}^{\tau/2}\hat{Q}_{\xi}(u)du \leq  \hat{Q}_{\xi}(\tau/2)\, ,$$
and
$$\hat{Q}_{\xi}(1 - p) \land \frac{1}{\tau/2}\int_{1-\tau/2}^{1}\hat{Q}_{\xi}(u)du \geq  \hat{Q}_{\xi}(1- \tau/2)\ , .$$

Next, the inequality $\operatorname{length}(\mathcal{I}_1) \geq \operatorname{length}(\mathcal{I}_2)$ follows immediately from the construction of Method 2. It remains to show that, in the symmetric case, we have that $\mathcal{I}_1 \supseteq \mathcal{I}_2$ and $\mathcal{I}_1 \supseteq \mathcal{I}_2$. This can be verified by noticing that, when the distribution of residuals is symmetric, we have that:

$$\hat{Q}_{|\xi|}(1-p) = \hat{Q}_{\xi}(1-p/2) = -\hat{Q}_{\xi}(p/2) \, . $$

        An immediate consequence of this fact is that the condition $2c^* < \hat U - \hat L$ underpinning the construction of \eqref{eq_refined} can be rewritten as:
        \begin{equation*}
            \begin{aligned}              \sum_{i=1}^{N_1}\frac{1}{N_1}h(Z_{(N_1+1-i)};\hat{\delta}) \hat{Q}_{\xi}\left(1-i\frac{\tau}{2N_1}\right) < \sum_{i=1}^{N_1}\frac{1}{N_1}h(Z_{i};\hat{\delta})  \min\left\{ \hat{Q}_{\xi}\left(1-\frac{\tau}{2N_1}\right), \frac{1}{\tau/2}\int_{1-\tau/2}^{1}\hat{Q}_{\xi}(u)du \right\}\, ,
            \end{aligned}
        \end{equation*}
     and, since under symmetry, one may write:
$$\mathcal{I}_1 = \hat{\alpha}\pm \sum_{i=1}^{N_1}\frac{1}{N_1}h(Z_{i};\hat{\delta})  \min\left\{ \hat{Q}_{\xi}\left(1-\frac{\tau}{2N_1}\right), \frac{1}{\tau/2}\int_{1-\tau/2}^{1}\hat{Q}_{\xi}(u)du \right\}\, ,$$
  and   $$\mathcal{I}_{BH}= \hat{\alpha}\pm \sum_{i=1}^{N_1}\frac{1}{N_1}h(Z_{(N_1+1-i)};\hat{\delta}) \hat{Q}_{\xi}\left(1-i\frac{\tau}{2N_1}\right) \, , $$
  we conclude that $\mathcal{I}_{2} \subseteq \mathcal{I}_1$. The inclusion $\mathcal{I}_{3} \subseteq \mathcal{I}_2$ then follows immediately from the fact that $\hat{Q}_{\xi}\left(1-i\frac{\tau}{2N_1}\right) \geq \hat{Q}_{\xi}\left(1-\frac{\tau}{2}\right) $, for any $i=1,\ldots, N_1$, and that, under symmetry:

  $$\mathcal{I}_3 = \hat{\alpha}\pm \sum_{i=1}^{N_1}\frac{1}{N_1}h(Z_{i};\hat{\delta}) \hat{Q}_{\xi}\left(1-\frac{\tau}{2N_1}\right)\, .$$
  \item \textbf{Weighted case.} The crucial step lies in showing that $\mathcal{I}_4 \subseteq \mathcal{I}_3$. The other relations can be established by applying the same arguments in the previous item to versions of the confidence intervals built upon the weighted DID estimator. We begin by noticing that, in its weighted version, the CI implied by inversion of Method 3 can be written as:
  
$$\mathcal{I}_3 = \left[\hat \alpha_\omega -\sum_{i=1}^{N_1} \omega_i \sqrt{\hat A + \frac{\hat{B}}{M_i}} \hat{Q}_{\xi}\left(1 - \frac{\tau}{2}\right), \hat \alpha_\omega - \sum_{i=1}^{N_1}\omega_i \sqrt{\hat A + \frac{\hat{B}}{M_i}} \hat{Q}_{\xi}\left(\frac{\tau}{2}\right) \right]\, ,$$
  where $\omega_i = \frac{M_i}{M_T}$ are the weights attached to cluster $i$, and $\hat \alpha_\omega$ denotes the weighted DID estimator. In contrast, the CI resulting from Method 4 is given by:
  
$$\mathcal{I}_4 = \left[\hat \alpha_\omega -\sqrt{\hat A + \frac{\hat{B}}{M_T}} \hat{Q}_{\xi}\left(1 - \frac{\tau}{2}\right), \hat \alpha_\omega -  \sqrt{\hat A + \frac{\hat{B}}{M_T}} \hat{Q}_{\xi}\left(\frac{\tau}{2}\right) \right]\, .$$

The inclusion then follows from observing that:

$$ \hat A + \frac{\hat{B}}{M_i} \geq \hat A + \frac{\hat{B}}{M_T}\quad  \forall i \implies \sum_{i=1}^{N_1}\omega_i \sqrt{\hat A + \frac{\hat{B}}{M_i}} \geq \sqrt{\hat A + \frac{\hat{B}}{M_T}}\, .  $$
    \end{enumerate}
\end{proof}

\subsection{Proof of Lemma \ref{Lemma_elliptical}}
\label{proof_elliptical}
    The proof of our result follows from the two claims below:

    \begin{claim}
    \label{claim_cf}
     $\boldsymbol{V}$ is a $k \times 1$ random vector whose joint distribution is invariant to orthogonal transformations if, and only if, its characteristic function  $\phi_{\boldsymbol{V}}$ is constant on the spheres $S_c = \{t \in \mathbb{R}^k : \lVert t \rVert_2 = c\}$, for any $c \geq 0$.
    \end{claim}
    \begin{proof}
    Suppose that the the characteristic function is constant on the spheres. Let $B \in \mathbb{R}^{k \times k}$ be an orthogonal matrix. It is not difficult to see that, for any $t \in \mathbb{R}^k$:

    $$\phi_{B\boldsymbol{V}}(t) = \mathbb{E}[\exp(it'B\boldsymbol{V})] = \phi_{\boldsymbol{V}}(B't) \overset{\lVert B' t\rVert_2 = \lVert t \rVert_2 }{=} \phi_{\boldsymbol{V}}(t)\, .$$
    
Thus showing that the characteristic function is invariant to orthogonal transformations. The desired conclusion then follows from the multivariate version of the inversion theorem (Corollary 4 of \cite{Shephard1991}).

Conversely, assume now that the distribution of $\boldsymbol{V}$ is invariant to orthogonal transformations. Consider $t, u \in \mathbb{R}^k$ with $t \neq u$ and $\lVert t \rVert  = \lVert u \rVert$. Put $B = \mathbb{I}_{k \times k} - 2zz'$, where $z = \frac{t - u}{\lVert t - u \rVert}$. It is not difficult to see that $B$ is orthogonal and that $Bt = u$. Consequently, we obtain that:

$$\phi_{\boldsymbol{V}}(t) = \mathbb{E}[\exp(i t'\boldsymbol{V})] \overset{\boldsymbol{V}\overset{d}{=}B'\boldsymbol{V}}{=} \mathbb{E}[\exp(i t'B'\boldsymbol{V})] = \phi_{\boldsymbol{V}}(Bt) =  \phi_{\boldsymbol{V}}(u)\, ,$$
thus proving the desired result.
\end{proof}

\begin{claim}
    \label{claim_invariance}
    If $\boldsymbol{V}$ is a $k \times 1$ random vector whose joint distribution is invariant to orthogonal transformations, then the entries of $\boldsymbol{V}$ are identically distributed, uncorrelated, have zero mean and are symmetric about zero.
\end{claim}
\begin{proof}
    Let $\pi$ be a permutation on $\{1,\ldots,k\}$, and $B = \begin{bmatrix}
        e_{\pi(1)} & e_{\pi(2)} & \ldots & e_{\pi(K)}
    \end{bmatrix}$ the associated permutation matrix, where $e_j$ is the $k \times 1$ vector with one in the $j$-th entry and zero otherwise. Since $B$ is orthogonal, $\boldsymbol{V} \overset{d}{=} B \boldsymbol{V}$. Since $\pi$ was arbitrarily chosen, we conclude that $\boldsymbol{V}$ is exchangeable and thus that its entries are identically distributed. Next, fix $i \in \{1,\ldots k\}$, and consider $B$ as the diagonal matrix with $B_{i,i}=-1$ and $B_{j,j}=1$ if $j \neq i$. Since this is an orthogonal matrix, we obtain that $\boldsymbol{V} \overset{d}{=} B \boldsymbol{V} \implies \mathbb{V}[\boldsymbol{V}] = \mathbb{V}[B\boldsymbol{V}]$, which yields that, for every $j \neq i$: $\operatorname{cov}(\boldsymbol{V}_i,\boldsymbol{V}_j) = \operatorname{cov}(-\boldsymbol{V}_i,\boldsymbol{V}_j) \implies \operatorname{cov}(\boldsymbol{V}_i,\boldsymbol{V}_j) = 0$. Moreover, $\boldsymbol{V} \overset{d}{=} B \boldsymbol{V} $ implies that $\mathbb{E}[\boldsymbol{V}_i] = \mathbb{E}[-\boldsymbol{V}_i] \implies \mathbb{E}[\boldsymbol{V}_i]=0$ and $\boldsymbol{V}_i \overset{d}{=}-\boldsymbol{V}_i$. Since $i$ was arbitrarily chosen, we conclude that the entries of $\boldsymbol{V}$ are uncorrelated, zero-mean and symmetrically distributed about zero, as desired.
\end{proof}

With the aforementioned facts in mind, we now prove Lemma \ref{Lemma_elliptical}.

\begin{proof}
    We first apply Claim \ref{claim_cf} to our context, from which we obtain that $\boldsymbol{\xi}:= (\xi_{i})_{i \in \mathcal{I}_1}$ has characteristic function $\phi_{\boldsymbol{\xi}}(t) = \varphi(t'AA't)$ for some function $\varphi: \mathbb{R} \mapsto \mathbb{C}$. Our desired conclusion then follows by observing that, for any linear combination $s \in \mathbb{R}^{|\mathcal{I}_1|}$, $\frac{s'\boldsymbol{\xi} }{\sqrt{\mathbb{V}[s'\boldsymbol{\xi} ]}} \overset{d}{=} \frac{\xi_1}{\sqrt{{\mathbb{V}[\xi_1]}}}$. This can be verified by, first, noticing that Claim \ref{claim_invariance} implies that, for some $\tau > 0$, $\mathbb{V}[\boldsymbol{\xi}] = \tau A A'$. Consequently, we obtain that $\mathbb{V}[s'\boldsymbol{\xi} ] = \tau s'AA's$ and $\mathbb{V}[\xi_1 ] = \tau e_1'AA'e_1$. Next, we observe that the characteristic functions:

$$\phi_{\frac{s'\boldsymbol{\xi} }{\sqrt{\mathbb{V}[s'\boldsymbol{\xi} ]}}}(\ell) = \phi_{\boldsymbol{\xi}}\left(\frac{\ell s}{\sqrt{\mathbb{V}[s'\boldsymbol{\xi}]}}  \right) = \varphi\left(\frac{\ell}{\tau}\right)\, ,$$
and
$$\phi_{\frac{\xi_1}{\sqrt{{\mathbb{V}[\xi_1]}}}}(\ell) = \phi_{\boldsymbol{\xi}}\left(\frac{\ell e_1}{\sqrt{\mathbb{V}[e_1'\boldsymbol{\xi}]}}  \right) = \varphi\left(\frac{\ell}{\tau}\right)\, ,$$
coincide for every $\ell \in \mathbb{R}$. The equality in distributions then follows from the inversion theorem. We thus have that linear combinations of $s'\boldsymbol{\xi}$ form a scale family with ``base'' quantile given by $Q_{\xi_1}$. Next, since $\xi_1$ is symmetric about zero, we have that $Q_{\xi_1}(1-u) \geq 0 \geq Q_{\xi_1}(u)$, for any $u \in (0,1/2)$. Consequently, we obtain that the most extreme scenario for the tail quantiles of a linear combination of the $s'\boldsymbol{\xi}$ with nonnegative weights $s\geq 0$ is attained when the variance $s'\boldsymbol{\xi}$ is the largest, i.e. at the comonotone copula.
\end{proof}

\subsection{Proof of Lemma \ref{lemma_uniform_qt}}
\label{proof_uniform_qt}

\begin{proof}
 Consider the case where $0<\underline{p} < \overline{p} < 1$. Fix $\epsilon > 0$. Since $F$ is continuous, $Q_F$ is continuous on $[\underline{p},\overline{p}]$, and thus uniformly continuous on $[\underline{p},\overline{p}]$. Consequently, for a given $\epsilon>0$, there exists a collection of finite closed intervals $([a_j,b_j])_{j=1}^k$, such that $\cup_{j=1}^k [a_j,b_j] = [\underline{p},\overline{p}]$ with the property that:

 $$x', x'' \in I_j \implies |Q_F(x')-Q_F(x'')|\leq \epsilon\, .$$

 Using the monotonicity of quantile functions, we thus obtain that:

  $$\sup_{p \in [\underline{p},\overline{p}]} |Q_{F_n}(p) - Q_F(p)| \leq \epsilon + \max_{j=1,\ldots, k}|Q_{F_n}(b_j) - Q_F(b_j)| + \max_{j=1,\ldots, k}|Q_{F_n}(b_j) - Q_{F_n}(a_j)| $$

 Now, since $F$ is continuous and $F_n(x)$ converges in probability to $F(x)$ for every $x \in \mathbb{R}$, $Q_{F_n}(x) \overset{p}{\to} Q_F(x)$ for every $x \in \mathbb{R}$ \citep[Lemma 21.2]{Vaart1998}. Consequently, by the continuous mapping theorem, $\max_{j=1,\ldots, k}|Q_{F_n}(b_j) - Q_F(b_j)|  \overset{p}{\to} 0$ and $\max_{j=1,\ldots, k}|Q_{F_n}(b_j) - Q_{F_n}(a_j)| \overset{p}{\to} \max_{j=1,\ldots, k}|Q_{F}(b_j) - Q_{F}(a_j)|\leq \epsilon  $. Now consider the event $A_n = \{\max_{j=1,\ldots, k}|Q_{F_n}(b_j) - Q_F(b_j)| > \epsilon \} \cup \{ ||Q_{F_n}(b_j) - Q_{F_n}(a_j)| - \max_{j=1,\ldots, k}|Q_{F}(b_j) - Q_{F}(a_j)||>\epsilon\}$. Since $\mathbb{P}[A_n] \to 0$, we have $P[A_n^\complement ] \to 1$. But then, using the preceding inequality, we note that $A_n^\complement \subseteq \{|Q_{F_n}(p) - Q_F(p)| \leq 4\epsilon\}$, from which it follows that:

   $$\mathbb{P}\left[\sup_{p \in [\underline{p},\overline{p}]} |Q_{F_n}(p) - Q_F(p)| \leq 4\epsilon \right] \to 1 \, .$$

   Finally, since the choice of $\epsilon$ was arbitrary, we obtain the desired result. The case with bounded support is similar, for in this setting $|Q(0)|\land |Q(1)| < \infty$, and we may thus repeat the argument with $0=\underline{p}<\overline{p}=1$. The cases of left- and right-bounded supports are analogous.
\end{proof}

\subsection{Proof of Lemma \ref{lemma_ES_CONVERGENCE}}
\label{proof_es_conv}
\begin{proof}
Fix $\delta > 0$ and consider some arbitrary $\varepsilon< \delta$.  Since $\mu^+_\tau < \infty$, one can find  $c\in(\tau,1)$ sufficiently close to $1$ so that
\begin{equation}\label{eq:tail-small}
  (1-c)\,|Q(c)| \leq    \int_c^1 |Q(u)|\,du < \varepsilon 
\end{equation}
and such that $F$ is continuous on $Q(c)$ (recall that $F$ has countably mane discontinuity points).
Decompose
\[
\int_{\tau}^1 |\hat Q_n(u)-Q(u)| du
\;\le\;
A_n(c) + B_n(c), \]
where
\[
A_n(c):=\int_{\tau}^c|\hat Q_n(u)-Q(u)| du, \quad
B_n(c):=\int_c^1|\hat Q_n(u)-Q(u)| du.
\]

We analyze each block separately.

\emph{(1) Interior block.}
For the fixed $c<1$ in \eqref{eq:tail-small},
\[
A_n(c)\le (c-\tfrac12)\sup_{u\in[\tau,c]}|\hat Q_n(u)-Q(u)|
\xrightarrow{p}0.
\]

\emph{(2) Tail block.}
By monotonicity of $\hat Q_n$ and $Q$, for $u\ge c$,
\[
|\hat Q_n(u)-Q(u)|
\le (\hat Q_n(u)-\hat Q_n(c)) 
    + (Q(u)-Q(c))
    + |\hat Q_n(c)-Q(c)|.
\]
Integrating over $[c,1]$ gives
\begin{equation}\label{eq:Bn-decomp}
B_n(c)
\le T_{n,1}+T_0+(1-c)|\hat Q_n(c)-Q(c)|,
\end{equation}
where
\[
T_{n,1}:=\int_c^1(\hat Q_n(u)-\hat Q_n(c))\,du,
\qquad
T_0:=\int_c^1(Q(u)-Q(c))\,du.
\]

The population term satisfies $T_0<2\varepsilon$ by \eqref{eq:tail-small}, and 
$(1-c)|\hat Q_n(c)-Q(c)|\xrightarrow{p}0$ by interior consistency.
For the first term, we observe that:
\[
T_{n,1}\le \frac{1}{n}\sum_{i=1}^n (X_i-\hat Q_n(c))_+ .
\]

Let $\eta>0$ be such that $\int_{F(Q(c)-\eta)}^1 |Q(u)|du < \epsilon$ and $\eta(1-F(Q(c)-\eta)) < \epsilon$.\footnote{One can find such $\eta$ due to continuity of $F$ at $Q(c)$.} Let $\mathcal G_{n,\eta}=\{|\hat Q_n(c)-Q(c)|\le\eta\}$. 
On $\mathcal G_{n,\eta}$,
\[
(X_i-\hat Q_n(c))_+ \le (X_i-(Q(c)-\eta))_+ .
\]

Hence:
\[
\mathbb P(T_{n,1}>4\delta)
\le \mathbb P(\mathcal G_{n,\eta}^c)
   + \mathbb P\!\left(\frac1n\sum_{i=1}^n (X_i-(Q(c)-\eta))_+ > 4\delta\right).
\]
The first term vanishes by (i) as $n \to \infty$. For the second, by Markov’s inequality,
\[
\mathbb P\!\left(\frac1n\sum_{i=1}^n (X_i-(Q(c)-\eta))_+ > 4\delta\right)
\le 
\frac{\mathbb E[(X_1-(Q(c)-\eta))_+]}{4\delta}.
\]
Using the quantile representation of $X_1$,
$\mathbb E[(X_1-t)_+]=\int_{F(t)}^1 (Q(u)-t)\,du$
with $t=Q(c)-\eta$. One then obtains that:
\[
\mathbb E[(X_1-(Q(c)-\eta))_+]
\le \int_{F(t)}^1 (Q(u)-Q(c))_+\,du + \eta(1-F(t)) < 3 \epsilon,
\]

Consequently, we obtain that:

$$\limsup_{n \to \infty} \mathbb{P}[T_{n,1} > 4\delta]< \frac{3}{4} \frac{\epsilon}{\delta}$$

Connecting each of the above steps, we obtain from the union bound that:

\begin{equation*}
    \begin{aligned}
        \mathbb{P}\left[\int_{\tau}^1 \!|\hat Q_n(u)-Q(u)|\,du > \delta\right] \leq \\ \mathbb{P}\left[A_n(c)> 4\delta\right] + \mathbb{P}\left[T_{n,1}> 4\delta\right] + \mathbb{P}\left[T_0> 4\delta \right]+\mathbb{P}\left[(1-c)|\hat Q_n(c)-Q(c)| > 4 \delta \right] 
    \end{aligned}\, ,
\end{equation*}
thus implying that:
$$ \limsup_{n \to \infty} \mathbb{P}\left[\int_{\tau}^1 \!|\hat Q_n(u)-Q(u)|\,du > \delta\right] \leq \frac{3}{4}\frac{\epsilon}{\delta}\, ,$$
and, since $\epsilon$ can be taken to be arbitrarily small, we conclude, that.

$$\lim_{n \to \infty} \mathbb{P}\left[\int_{\tau}^1 \!|\hat Q_n(u)-Q(u)|\,du > \delta\right] = 0 \, ,$$
thus establishing the desired result.
\end{proof}


\singlespacing


\bibliographystyle{apalike}
\bibliography{bib.bib}